\documentclass[longauth]{aa}

\usepackage[T1]{fontenc}
\usepackage{ae,aecompl}
\usepackage{graphicx}   
\usepackage{txfonts}    
\usepackage{amssymb}    
\usepackage{xspace}
\usepackage[table]{xcolor}
\usepackage{scalerel}
\usepackage{tikz}
\usepackage{newtxtext,newtxmath}
\usepackage{hyperref}
\usepackage{caption}

\newcommand{\angstrom}{\,\textup{\AA}\xspace}
\newcommand{\Msun}{\ensuremath{\mathrm{M}_\odot}\xspace}

\newcommand{\kms}{\ensuremath{\mathrm{km\,s}^{-1}}\xspace}
\newcommand{\ergs}{\ensuremath{\mathrm{erg\,s}^{-1}}\xspace}

\newcommand{\hbeta}{\ensuremath{\mathrm{H}\beta}\xspace}
\newcommand{\mgii}{\ion{Mg}{ii}\xspace}
\newcommand{\civ}{\ion{C}{iv}\xspace}

\begin{document} 

   \title{Black hole virial masses from single-epoch photometry}
   \authorrunning{J. Chaves-Montero et al.}

   \subtitle{The miniJPAS test case}
    
    \author{
    J.~Chaves-Montero\inst{1}\fnmsep\thanks{E-mail: \href{mailto:jonas.chaves@dipc.org}{jonas.chaves@dipc.org}}, 
    S.~Bonoli\inst{1,2}, 
    B.~Trakhtenbrot\inst{3}, 
    A.~Fern\'{a}ndez-Centeno\inst{4}, 
    C.~Queiroz\inst{5,6},
    L.~A.~D\'iaz-Garc\'ia\inst{7},
    R.~M.~Gonz\'alez Delgado\inst{7},
    A.~Hern\'an-Caballero\inst{8},
    C.~Hern\'andez-Monteagudo\inst{9,10},
    C.~L\'open-Sanjuan\inst{11},
    R.~Overzier\inst{12,13},
    D.~Sobral\inst{14},
    L.~R. Abramo\inst{6},
    J.~Alcaniz\inst{5,12},
    N.~Benitez\inst{7},
    S.~Carneiro\inst{15},
    A.~J. Cenarro\inst{11},
    D.~Crist\'obal-Hornillos\inst{8},
    R.~A. Dupke\inst{12,16,17},
    A.~Ederoclite\inst{8},
    A.~Mar\'in-Franch\inst{11},
    C.~Mendes~de~Oliveira\inst{13},
    M.~Moles\inst{7,8},
    L.~Sodr\'e~Jr.\inst{13},
    K.~Taylor\inst{18},
    J.~Varela\inst{11},
    H.~V\'azquez~Rami\'o\inst{11},
    \and T.~Civera\inst{8}
    }

   \institute{
    Donostia International Physics Center, Paseo Manuel de Lardizabal 4, E-20018 Donostia-San Sebastian, Spain.
    \goodbreak
    \and
    Ikerbasque, Basque Foundation for Science, E-48013 Bilbao, Spain. 
    \goodbreak
    \and
    School of Physics and Astronomy, Tel Aviv University, Tel Aviv 69978, Israel.
    \goodbreak
    \and
    Facultad de Ciencias F\'{i}sicas, Universidad Complutense de Madrid, Plaza de Ciencias, 1, 28040 Madrid, Spain.
    \goodbreak
    \and
    Departamento de Astronomia, Instituto de F\'isica, Universidade Federal do Rio Grande do Sul (UFRGS), Av. Bento Gon\c{c}alves 9500, Porto Alegre, RS, Brazil.
    \goodbreak
    \and
    Instituto de F\'isica, Universidade de S\~ao Paulo, Rua do Mat\~ao 1371, CEP 05508-090, S\~ao Paulo, Brazil.
    \goodbreak
    \and
    Instituto de Astrof\'isica de Andaluc\'ia (CSIC), P.O.~Box 3004, 18080 Granada, Spain. 
    \goodbreak
    \and
    Centro de Estudios de F\'{i}sica del Cosmos de Arag\'on (CEFCA), Plaza San Juan, 1, E-44001 Teruel, Spain.
    \goodbreak
    \and
    Instituto de Astrof\'isica de Canarias, Calle V\'ia L\'actea SN, ES38205 La Laguna, Spain.
    \goodbreak
    \and
    Departamento de Astrof\'isica, Universidad de La Laguna, ES38205 La Laguna, Spain.
    \goodbreak
    \and
    Centro de Estudios de F\'{i}sica del Cosmos de Arag\'on (CEFCA), Unidad Asociada al CSIC, Plaza San Juan, 1, E-44001 Teruel, Spain.
    \goodbreak
    \and
    Observat\'{o}rio Nacional/MCTIC, Rua General Jos\'{e} Cristino, 77, S\~{a}o Crist\'{o}v\~{a}o, Rio de Janeiro, RJ 20921-400, Brazil.
    \goodbreak
    \and
    Institute of Astronomy, Geophysics and Atmospheric Sciences, University of S\~{a}o Paulo, Rua do Mat\~{a}o, 1226, S\~{a}o Paulo, SP 05508-090, Brazil.
    \goodbreak
    \and
    Department of Physics, Lancaster University, Lancaster, LA1 4YB, UK.
    \goodbreak
    \and
    Instituto de F\'isica, Universidade Federal da Bahia, 40210-340, Salvador, BA, Brazil.
    \goodbreak
    \and
    Department of Astronomy, University of Michigan, 311 West Hall, 1085 South University Ave., Ann Arbor, USA.
    \goodbreak
    \and
    University of Alabama, Department of Physics and Astronomy, Gallalee Hall, Tuscaloosa, AL 35401, USA.
    \goodbreak
    \and
    Instruments4, 4121 Pembury Place, La Canada Flintridge, CA 91011, USA.
    }
    
  \date{Received November 2, 2021; accepted January 28, 2022}

  \abstract
   {Precise measurements of black hole masses are essential to understanding the coevolution of these sources and their host galaxies.}
   {We develop a novel approach for computing black hole virial masses using measurements of continuum luminosities and emission line widths from partially overlapping, narrow-band observations of quasars; we refer to this technique as single-epoch photometry.}
   {This novel method relies on forward-modelling quasar observations for estimating emission line widths, which enables unbiased measurements even for lines coarsely resolved by narrow-band data. We assess the performance of this technique using quasars from the Sloan Digital Sky Survey (SDSS) observed by the miniJPAS survey, a proof-of-concept project of the Javalambre Physics of the Accelerating Universe Astrophysical Survey (J--PAS) collaboration covering $\simeq1\,\mathrm{deg}^2$ of the northern sky using the 56 J--PAS narrow-band filters.}
   {We find remarkable agreement between black hole masses from single-epoch SDSS spectra and single-epoch miniJPAS photometry, with no systematic difference between these and a scatter ranging from 0.4 to 0.07 dex for masses from $\log(M_\mathrm{BH})\simeq8$ to 9.75, respectively. Reverberation mapping studies show that single-epoch masses present approximately 0.4 dex precision, letting us conclude that our novel technique delivers black hole masses with only mildly lower precision than single-epoch spectroscopy.}
   {The J--PAS survey will soon start observing thousands of square degrees without any source preselection other than the photometric depth in the detection band, and thus single-epoch photometry has the potential to provide details on the physical properties of quasar populations that do not satisfy the preselection criteria of previous spectroscopic surveys.}

   \keywords{quasars: supermassive black holes -- 
             quasars: emission lines -- 
             galaxies: photometry -- 
             galaxies: active -- 
             line: profiles
               }

   \maketitle
%

\section{Introduction}

Quasars are the most luminous persistent sources known; as a result, they enable us to study the Universe from late to very early epochs \citep[e.g.][]{Fan2006, mortlock11, banados2018_800millionsolarmassBlackhole, yang2020_PoniuaEnaLuminous, wang2021_LuminousQuasarRedshift}. For example, quasars are excellent large-scale structure tracers at redshifts where the number density of bright galaxies is too low for statistical studies \citep[e.g.][]{busca2013_BaryonAcousticoscillations, castorina2019_RedshiftweightedConstraintsprimordial, hou2021_CompletedSDSSIVextended}, provide crucial information about the re-ionisation history of the Universe \citep[e.g.][]{miralda-escude1998_ReionizationIntergalacticMedium, Fan2006, banados2018_800millionsolarmassBlackhole, davies2018_QuantitativeConstraintsReionization, wang2020_SignificantlyNeutralIntergalactic, yang2020_MeasurementsIntergalacticMedium}, and have been proposed as standardisable candles \citep[e.g.][]{watson11, wang14, risaliti2019_CosmologicalConstraintsHubble}, which provides a new avenue to extending Hubble parameter constraints towards high redshift.

The accepted physical picture is that a quasar is powered by the accretion of matter onto a supermassive black hole \citep[SMBH; e.g.][]{hoyle1963_NatureStrongradio, salpeter1964_AccretionInterstellarMatter, lynden-bell1969_GalacticNucleiCollapsed}, which is inferred to exist at the centre of every massive galaxy \citep[e.g.][]{kormendy95, magorrian1998_DemographyMassiveDark, ferrarese2000_FundamentalRelationSupermassive}. The discovery of correlations between multiple galaxy properties and SMBH mass \citep[e.g.][]{magorrian1998_DemographyMassiveDark, ferrarese2000_FundamentalRelationSupermassive, gebhardt00, gultekin2009_MsMLRelations, kormendy2013_CoevolutionNotSupermassive, mcconnell2013_RevisitingScalingRelations} suggests a coevolution between SMBHs and their host galaxies, during which the energy released by the accreting SMBH self-regulates its growth and impacts the evolution of its host \citep{silk1998_QuasarsGalaxyformation, king2003_BlackHolesGalaxy, dimatteo2005_EnergyInputquasars}. The coevolution scenario is also supported by the similar redshift evolution of the cosmic SMBH accretion rate and the cosmic star formation rate up to $z=4$ \citep{merloni2004_TracingCosmologicalassembly, silverman2008_LuminosityFunctionXRayselected, shankar2009_SelfConsistentModelsAGN, aird2010_EvolutionHardXray, delvecchio2014_TracingCosmicgrowth, yang2018_LinkingBlackhole} as well as the need for active galactic nucleus feedback to explain the stellar-to-halo mass relation for massive galaxies in hydrodynamical simulations and semi-analytic models \citep[e.g.][]{croton2006_ManyLivesactive, somerville2008_SemianalyticModelcoevolution, dubois2012_SelfregulatedGrowthsupermassive, sijacki2015_IllustrisSimulationevolving, schaye15, Weinberger2017, Pillepich2018a}. However, the nature of the SMBH-galaxy relation is still not completely understood; for instance, the presence of SMBHs with masses larger than $10^9\,\Msun$ at $z>6$ raises an important question about the origin and fast growth of these objects \citep[see][for a recent review]{inayoshi2020_AssemblyFirstMassive}.

In the local universe, SMBH masses are estimated by resolving the dynamics of stars \citep[e.g.][]{tremaine2002_SlopeBlackHole, marconi2003_RelationBlackHole, davies2006_StarformingTorusStellar, onken2007_BlackHoleMass} or gas \citep[e.g.][]{davies2004_NuclearGasDynamics, davies2004_NuclearGasdynamicsStar,hicks2008_CircumnuclearGasSeyfert} within the SMBH's gravitational sphere of influence. At cosmological distances, spatially resolving this region is impossible, and the standard approach to estimating SMBH masses for distant galaxies relies on measurements of the virial motion of gas in the broad-line region \citep[BLR; e.g.][]{czerny2011_OriginBroadline}. The most precise method for computing virial masses is reverberation mapping \citep[RM; e.g.][]{blandford1982_ReverberationMappingemission, peterson1993_ReverberationMappingActive, netzer1997_ReverberationMappingPhysics}, which yields robust SMBH mass estimates consistent with dynamical masses \citep[e.g.][]{bentz2013_LowluminosityEndRadiusLuminosity}. Reverberation mapping measures the velocity of clouds in the BLR from the width of broad emission lines and the BLR size from the time lag between the continuum and emission line variability; to do so, it requires multiple spectroscopic observations over an extended period of time at high cadence \citep[e.g.][]{macleod2010_ModelingTimeVariability, bentz2013_LowluminosityEndRadiusLuminosity}, which has limited the application of this technique to a few hundred sources so far \citep{kaspi2021_TakingLongLook}. 

The only method for estimating SMBH virial masses for a large number of sources is single-epoch spectroscopy \citep[SES; e.g.][]{wandel1999_CentralMassesBroadLine, mclure2002_MeasuringBlackhole, vestergaard2002_DeterminingCentralBlack}, which relies on the tight correlation between quasar continuum luminosity and BLR size \citep[e.g.][]{kaspi2000_ReverberationMeasurements17, kaspi2005_RelationshipLuminosityBroadLine, bentz2006_RadiusLuminosityRelationshipActive, bentz2009_RadiusLuminosityRelationshipActive, bentz2013_LowluminosityEndRadiusLuminosity, lira2018_ReverberationMappingLuminous} to compute SMBH masses from a single spectrum. These lesser requirements translate into noisier SMBH mass estimates compared to RM, which require empirical calibration either from RM \citep[e.g.][]{kaspi2000_ReverberationMeasurements17, kaspi2005_RelationshipLuminosityBroadLine, bentz2013_LowluminosityEndRadiusLuminosity} or internally based on the availability of multiple emission lines for the same object \citep[e.g.][]{mclure2002_MeasuringBlackhole, vestergaard2006_DeterminingCentralBlack, shen2011_CatalogQuasarProperties}. Taken together with the systematics involved in the measurement of line widths, these sources of uncertainty result in differences between SMBH masses from SES and RM as large as 0.5 dex \citep[e.g.][]{mclure2002_MeasuringBlackhole, vestergaard2006_DeterminingCentralBlack, shen2013_MassQuasars, peterson2014_MeasuringMassesSupermassivea}.

Traditionally, measuring SMBH masses was a prerogative of spectroscopic surveys because the spectral resolution of photometric surveys was too coarse to resolve even the broadest quasar emission lines, which present widths of thousands of km/s. In addition, photometric redshifts from broadband photometry do not present enough precision for unambiguous line identification. The emergence of medium- and narrow-band photometric surveys continuously covering a large wavelength range, such as the Subaru Cosmic Evolution Survey 20 \citep[Subaru COSMOS 20;][]{taniguchi2015_SubaruCOSMOS20, sobral2018_SlicingCOSMOSSC4K}, the Advance Large Homogeneous Area Medium Band Redshift Astronomical (ALHAMBRA) survey \citep{moles08}, the National Optical Astronomy Observatory (NOAO) Extremely Wide-Field Infrared Imager (NEWFIRM) Medium-Band Survey \citep[NMBS;][]{vandokkum2009_NEWFIRMMediumBandSurvey}, the Survey for High-z Absorption Red and Dead Sources \citep[SHARDS;][]{perez-gonzalez2013_SHARDSOpticalSpectrophotometric}, the Physics of the Accelerating Universe Survey \citep[PAUS;][]{eriksen2019_PAUSurveyearly}, and the Javalambre-Physics of the Accelerating Universe Astrophysical Survey \citep[J--PAS;][]{benitez2014_JPASJavalambrePhysicsAccelerated}, is progressively changing this picture. Multi-band photometric surveys have reached high enough spectral resolution to first detect broad emission lines \citep{chaves-montero2017_ELDARNewmethoda, lumbreras-calle2019_StarformingGalaxieslowredshift} and then detect narrow lines and approximately resolve the profile of broad lines \citep{alarcon2021_PAUSurveyimproved, bonoli2021_MiniJPASSurveypreview, martinez-solaeche2021_JPASMeasuringemission}.

In this work, we develop the first method for measuring SMBH virial masses from narrow-band `photospectra', photometric observations from a contiguous set of partially overlapping narrow-band filters. This technique estimates the virial velocity of BLR clouds from the width of broad emission lines and the size of the BLR from the continuum luminosity; given the similarity of this technique with SES, we dub this approach single-epoch photometry (SEP). We show that the resolution of J--PAS photospectra is too coarse for backward-modelling\footnote{Backward-modelling refers to the process of measuring some target property directly from observations, while forward-modelling indicates the process of first producing plausible values of such a property using a theoretical model, and then measuring it by comparing these values with observations.} emission line widths in an unbiased fashion; motivated by this, we combine forward-modelling quasar observations and Bayesian inference to measure continuum luminosities and emission line widths. To validate our methodology, we use 54 Sloan Digital Sky Survey \citep[SDSS;][]{york00} quasars observed by the miniJPAS survey \citep{bonoli2021_MiniJPASSurveypreview}, a proof-of-concept project of the J--PAS collaboration. By comparing SES masses from SDSS and SEP masses from miniJPAS, we find that SEP delivers unbiased SMBH mass estimates with only slightly less precision than SES measurements for most masses. Our findings open up the possibility of studying the physical properties of quasar populations that do not satisfy the preselection criteria of previous spectroscopic surveys but will be observed by future narrow-band surveys such as J--PAS.

The paper is organised as follows. In Sect. \ref{sec:data} we introduce the dataset that we use to calibrate the performance SEP. In Sect. \ref{sec:model} we describe our approach to measuring continuum luminosities, emission line properties, and SMBH virial masses from narrow-band data, and in Sect. \ref{sec:results} we use SDSS quasars observed by the miniJPAS survey to estimate the precision of the previous measurements. In Sect. \ref{sec:conclusions} we summarise our main findings and conclude.

Throughout this paper we consider {\it Planck} 2015 cosmological parameters \citep{planck14b}: $\Omega_{\rm m}= 0.314$, $\Omega_\Lambda = 0.686$, $\Omega_{\rm b} = 0.049$, $\sigma_8 = 0.83$, $h = 0.67$, and $n_{\rm s} = 0.96$. We use the term quasar to refer to unobscured active galactic nuclei with at least one emission line broader than $1000\,\kms$. Emission lines with central wavelengths smaller and larger than $\lambda=2000\angstrom$ are provided in vacuum and air wavelengths, respectively. All magnitudes are reported in the AB system \citep{oke1983_SecondaryStandardstars}. We use the symbol $\log$ to indicate decimal logarithms.


\section{Data}
\label{sec:data}


\subsection{Narrow-band data: miniJPAS}
\label{sec:data_minijpas}

The miniJPAS survey \citep{bonoli2021_MiniJPASSurveypreview} observed $\simeq1\,\mathrm{deg}^2$ of the northern sky using the J--PAS filter system, which includes $54$ partially overlapping narrow-band filters of full width at half maximum (FWHM) $\simeq145\angstrom$ covering the optical range from 3780 to $9100\angstrom$ and $2$ broader filters expanding over the UV and the near-infrared up to approximately 3100 and $10\,000\angstrom$, respectively. The observations were carried out using an interim camera mounted on the $2.5\,\mathrm{m}$ diameter Javalambre Survey Telescope at the Astrophysical Observatory of Javalambre, which will be the same telescope conducting observations for the J--PAS survey. This survey was designed to serve as a proof-of-concept for the J--PAS project \citep{benitez2014_JPASJavalambrePhysicsAccelerated}.

The footprint of miniJPAS covers the Extended Groth Strip (EGS) field partially, where ancillary data from the All-Wavelength EGS International Survey \citep[AEGIS;][]{davis2007} and SDSS \citep{york00} are publicly available. To facilitate the comparison with other surveys, each pointing of miniJPAS was observed not only with all J--PAS filters, but also with the broadband filters $u$, $g$, $r$, and $i$. The depth of miniJPAS in a circular aperture of 3\arcsec\ diameter reaches $m\simeq22-23.5\,\rm{AB}$ at $5\sigma$ for the 54 narrow-band filters and up to $m=24\,\rm{AB}$ for the broader filters. The primary catalogue of this survey contains more than $64\,000$ sources detected in the $r$ band with matched forced photometry in all other bands \citep[see][for more details]{bonoli2021_MiniJPASSurveypreview}.

The J--PAS filter system\footnote{\url{http://svo2.cab.inta-csic.es/svo/theory/fps3/index.php?mode=browse&gname=OAJ&asttype=}} \citep{brauneck2018_CustomizedBroadbandSloanfilters, brauneck2018_DenseGridnarrow} was designed to provide accurate photometric redshifts for both blue and red galaxies up to $z\sim1$ \citep{benitez09a, benitez2014_JPASJavalambrePhysicsAccelerated}, and for quasars up to $z\simeq6$ \citep{abramo12, chaves-montero2017_ELDARNewmethoda}. The first results from miniJPAS confirmed the expectations of sub-percent photo-$z$ precision \citep{bonoli2021_MiniJPASSurveypreview, hernan-caballero2021_MiniJPASSurveyphotometric}, the potential of the J--PAS filter system to detect and characterise emission line sources \citep{bonoli2021_MiniJPASSurveypreview, gonzalezdelgado2021_MiniJPASSurveyIdentification, martinez-solaeche2021_JPASMeasuringemission}, and more specifically to capture the main features of low redshift quasars \citep{2021IAUS..356...12B} using {\sc qsfit} \citep{calderone2017}. Furthermore, the William Herschel Telescope (WHT) Enhanced Area Velocity Explorer (WEAVE) Quasi-Stellar Object (WEAVE-QSO) survey \citep{pieri2016} will follow up with high spectral resolution $\sim400k$ J--PAS quasars at $z>2$, allowing our approach to be tested and calibrated further.


\subsection{Spectroscopic data: SDSS}
\label{sec:data_sdss}

The SDSS survey \citep{york00} also observed the EGS field, and thus we can use quasars with SES measurements from SDSS to estimate the performance of SEP for miniJPAS. In this section we describe the main characteristics of the SDSS data we use.

To validate our methodology, we use publicly available SES measurements from the 14th data release of the SDSS quasar value-added catalogue \citep[SDSS14Q;][]{rakshit2020_SpectralPropertiesquasars}, which contains 526\,356 sources observed by any of the stages of the SDSS survey up to and including this data release \citep{york00, eisenstein11, dawson2013_BaryonOscillationSpectroscopica, dawson16}. Quasars included in this catalogue satisfy two selection criteria: $i$-band absolute magnitude brighter than $M_i(z=2)=-20.5$ and at least one emission line broader than $\mathrm{FWHM}=500\,\kms$. For each source, the SDSS14Q catalogue includes the most robust spectroscopic redshift solution from SDSS \citep[see][]{paris2018_SloanDigitalSky}, the FWHM and equivalent width (EW) of the broadest emission lines,
the monochromatic continuum luminosity nearby these lines, and SMBH virial mass estimates based on these key spectroscopic measurements. The spectral information was measured using the publicly available multi-component spectral fitting code {\sc pyqsofit}\footnote{\url{https://github.com/legolason/PyQSOFit/}} \citep{guo2018_PyQSOFitPythoncode}, which uses multiple components to model the continuum emission and emission lines of each quasar separately  \citep[for a detailed description of the code and its applications see][]{guo2019_ConstrainingSubparsecbinary, shen2019_SloanDigitalSky}.

\begin{figure}
    \centering
    \includegraphics[width=0.495\textwidth]{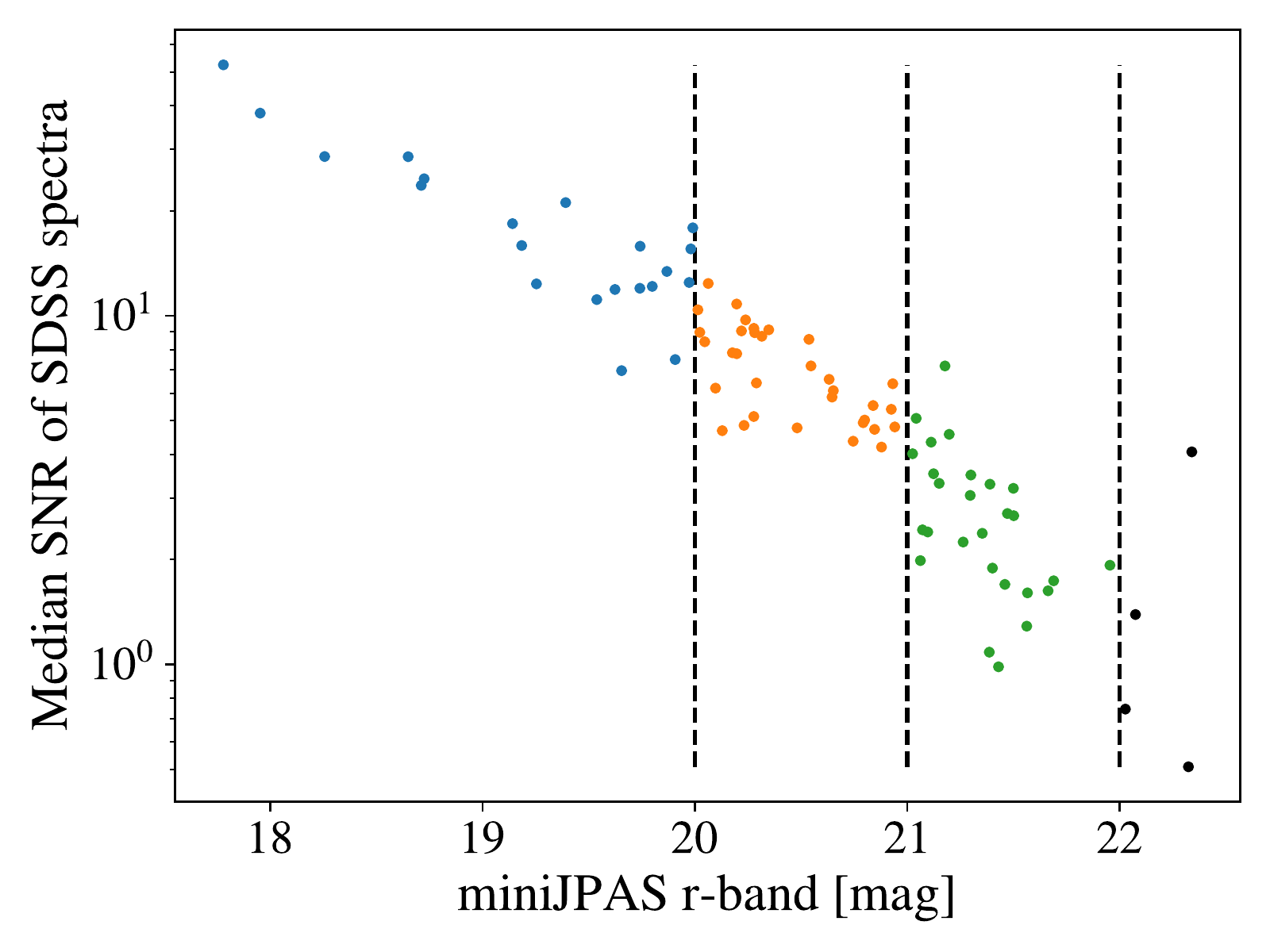}    \includegraphics[width=0.495\textwidth]{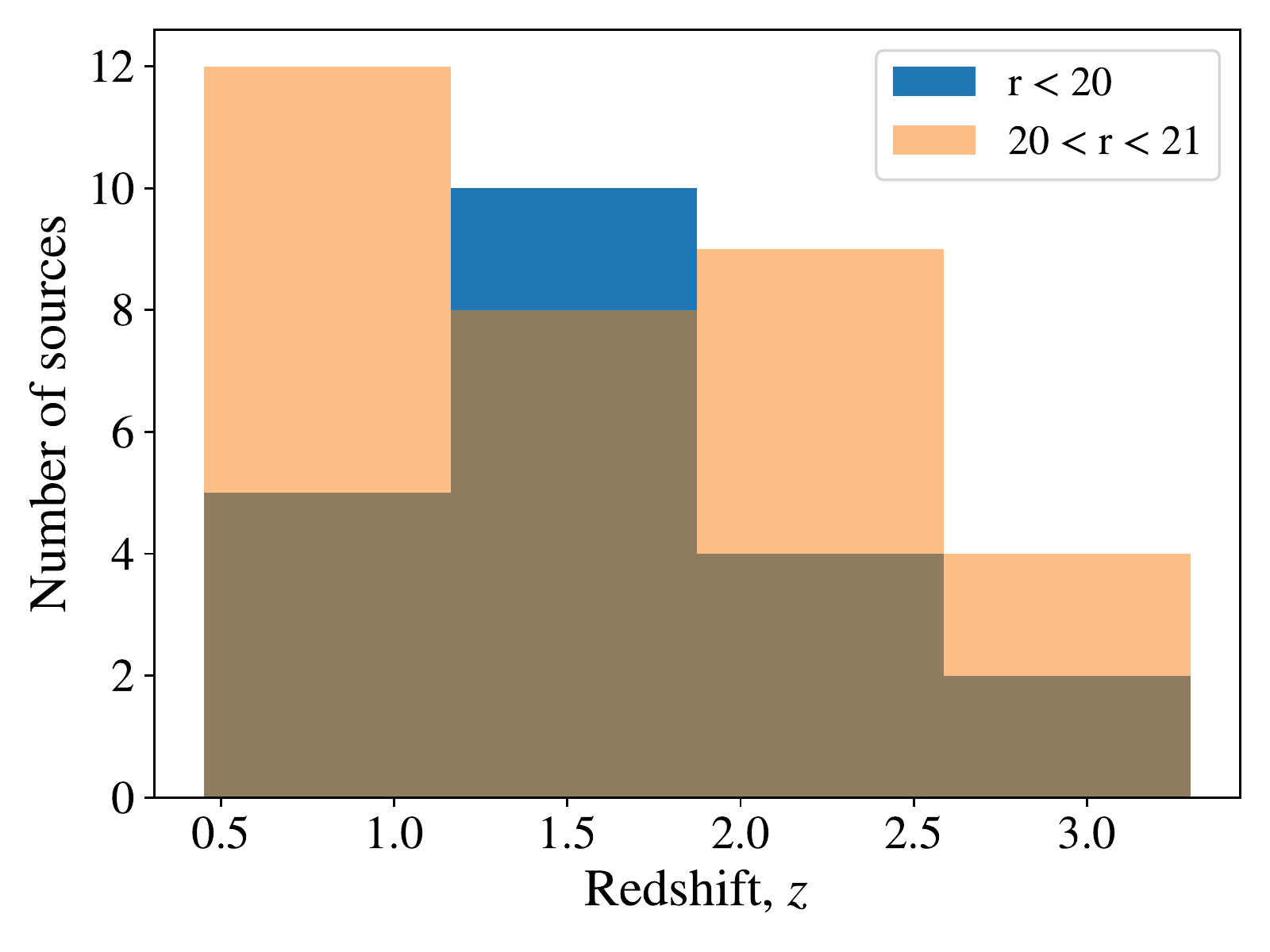}
    \caption{Properties of SDSS quasars with successful SES measurements observed by miniJPAS. Blue, orange, and green colours indicate the results for sources with $r<20$, $20<r<21$, and $21<r<22$, respectively. The top panel shows the median S/N of SDSS spectra as a function of miniJPAS $r$-band magnitude, and the bottom panel displays the redshift distribution of quasars brighter than $r=21$.}
    \label{fig:snr}
\end{figure}

\subsection{Validation sample}
\label{sec:data_sample}

We generate the quasar validation sample by cross-matching miniJPAS observations and the 12th data release of the SDSS quasar superset catalogue \citep[SDSS12Q,][]{paris17}, which contains visually inspected spectra and redshifts from any of the stages of the SDSS survey up to and including this data release. We find that miniJPAS observed 117 SDSS quasars and that 85 out of these present successful SES measurements from the SDSS14Q catalogue. 

We generate a photospectrum for each source by combining 3\arcsec\ aperture magnitudes from each of the miniJPAS narrow-band filters. We use this type of magnitude due to the point-like nature of quasars, and we correct aperture to total magnitudes following a two-step procedure. First, we compute the median difference between 3\arcsec\ aperture magnitudes of bright, unsaturated stars from each miniJPAS tile and point spread function magnitudes from the Panoramic Survey Telescope and Rapid Response System 1 \citep[Pan-STARRS1;][]{chambers2016_PanSTARRS1Surveys}. We use these offsets to correct the magnitudes of each tile separately, and then we compute the median difference between the resulting magnitudes and synthetic magnitudes obtained by convolving the spectra of 115 stars from the SDSS12Q catalogue with the J--PAS filter system. Finally, we apply these differences to the partially corrected magnitudes. We note that this two-step approach corrects for both the finite size of 3\arcsec\ apertures and spectral offsets (for more details about this process, see Queiroz et al. in prep.).

The miniJPAS survey conducted observations of most filters between May and October of 2018, and of a few filters in July 2019 \citep{bonoli2021_MiniJPASSurveypreview}. Due to the variable nature of quasars, we could expect variability to manifest as artificial emission or absorption lines in miniJPAS photospectra due to filters observed at different epochs. The impact of variability is increasingly weak for more luminous quasars \citep[e.g.][]{hook1994_VariabilityOpticallyselected, macleod2012_DescriptionQuasarVariability, meusinger2013_UltravioletVariabilityquasars, kozlowski2016_QuasarVariabilityMidInfrared, caplar2017_OpticalVariabilityAGNs}, and the expected level of optical variability is $\sim0.1\,\mathrm{AB}$ per 100 rest-frame days for the faintest sources. The maximum time span between miniJPAS observations is approximately 400 days, so we expect the largest band-to-band magnitude fluctuations to be smaller than 0.2 and 0.1 AB for quasars at $z=1$ and 3, respectively. Taken together with the high luminosity of miniJPAS quasars (see Sect. \ref{sec:results_continuum}), we expect minimal impact of variability on miniJPAS observations. On the other hand, the difference between SDSS and miniJPAS observations is of the order of years for some sources, and thus we expect variability to affect the comparison between SES and SEP measurements. We note that the virial theorem suggests that a change of $X$ dex in continuum luminosity manifests as a $-0.25\,X$ dex difference in FWHM.

In the top panel of Fig.~\ref{fig:snr}, we display the median spectral signal-to-noise ratio (S/N) of the 85 SDSS quasars with successful SES measurements as a function of their miniJPAS $r$-band magnitude. Blue, orange, and green dots indicate the results for sources with $r<20$, $20<r<21$, and $21<r<22$, respectively. As we can see, most sources with $r$-band magnitude fainter than $r=21$ present an SDSS spectrum with median S/N smaller than 4. Multiple authors have investigated the impact of S/N on the robustness of SES measurements \citep[e.g.][]{shen2011_CatalogQuasarProperties, denney2016_SloanDigitalSky, shen2019_SloanDigitalSky}, finding that the precision of SES masses decreases rapidly with the median S/N of SDSS spectra, reaching measurement-related errors of 0.3 dex or larger for $\mathrm{S/N}<10$ \citep[e.g.][]{shen2011_CatalogQuasarProperties, rakshit2020_SpectralPropertiesquasars}. To reduce the impact of noisy SES measurements on our analysis, in Sect. \ref{sec:results} we validate SEP using the 54 sources brighter than $r=21$.

In the bottom panel of Fig.~\ref{fig:snr}, we display the redshift distribution of these 54 quasars. As we can see, these sources present spectroscopic redshifts between $z\simeq0.5$ and 3.5, which enables SEP to be tested using the lines \hbeta, \mgii, and \civ. The maximum redshift for quasar detection in J--PAS is $z\simeq6$; nonetheless, the validation sample does not present any source above $z=4$ because the miniJPAS survey only observed $\simeq1\deg^2$ and the angular number density of quasars brighter than $r=21$ at $z>4$ is smaller than one per square degree \citep{palanque-delabrouille2013_LuminosityFunctiondedicated, palanque-delabrouille2016_ExtendedBaryonOscillation}. Throughout this work, we use SDSS redshift estimates to conduct SEP measurements; however, we will not have access to spectroscopic redshifts for the majority of sources that the J--PAS survey will observe. We expect photometric redshifts with subpercent precision for J--PAS quasars \citep[][]{abramo12, chaves-montero2017_ELDARNewmethoda, bonoli2021_MiniJPASSurveypreview}; as a result, photometric redshift errors are expected to be a subdominant source of uncertainty for our technique (see Appendix~\ref{app:errors}).


\section{Model}
\label{sec:model}

In this section we describe our novel approach to measure SMBH masses from SEP. We first discuss the theoretical foundations of this method in Sect. \ref{sec:model_theory}, and then we describe our strategy to measure continuum luminosities, emission line properties, and SMBH masses in Sects. \ref{sec:model_continuum}, \ref{sec:model_lines}, and \ref{sec:model_masses}, respectively. Lastly, we test our methodology using simulated J--PAS photospectra in Sect. \ref{sec:model_mock}.


\subsection{Theory preambles}
\label{sec:model_theory}

The standard approach to measure SMBH masses at cosmological distances relies on measurements of the virial motion of gas in the BLR. Assuming that the SMBH's gravitational field dominates the motion of these clouds, we can compute SMBH masses using the virial theorem \citep[e.g.][]{ho1999_SupermassiveBlackHoles, wandel1999_CentralMassesBroadLine}:
\begin{equation}
    M_\mathrm{BH}=f\frac{R_\mathrm{BLR}(\Delta V)^2}{G},
\end{equation}
where $G$ is the gravitational constant, $R_\mathrm{BLR}$ indicates the size of the BLR, $\Delta V$ refers to the virial velocity of the BLR gas, and $f$ is a dimensionless parameter of order unity that depends on the geometry, kinematics, and inclination of the BLR \citep[see][and references therein]{mejia-restrepo2018_EffectNucleargas, williams2018_LickAGNMonitoring}. In practice, it is standard to estimate the virial velocity using either the FWHM or the dispersion of broad emission lines, each of which presents different advantages and disadvantages \citep[e.g.][]{shen2013_MassQuasars}. Throughout the remainder of this section we describe a new method to measure SMBH masses from J--PAS photospectra by leveraging the tight correlation between continuum luminosity and BLR size.


\begin{figure}
    \centering
    \includegraphics[width=\columnwidth]{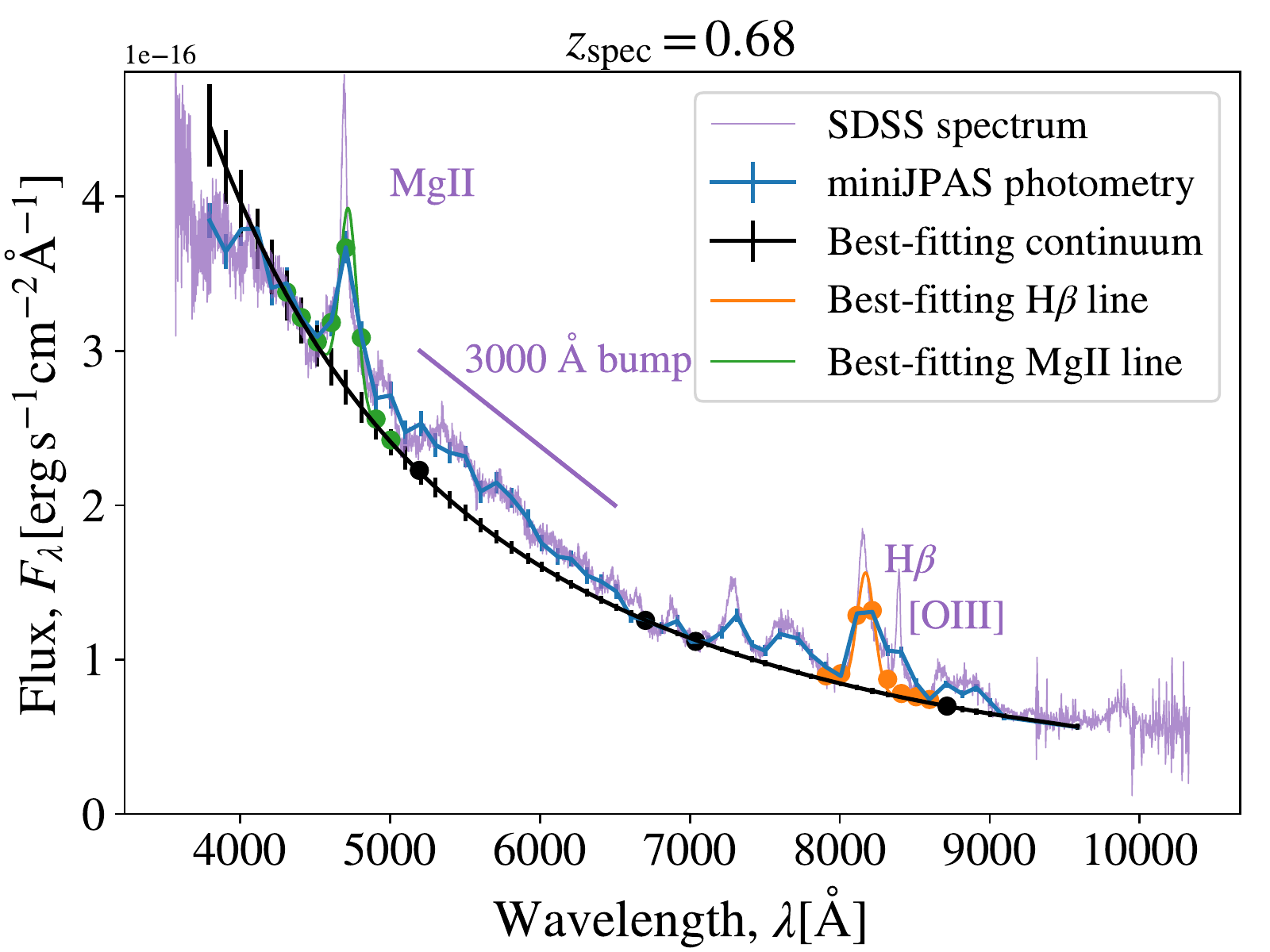}
    \includegraphics[width=\columnwidth]{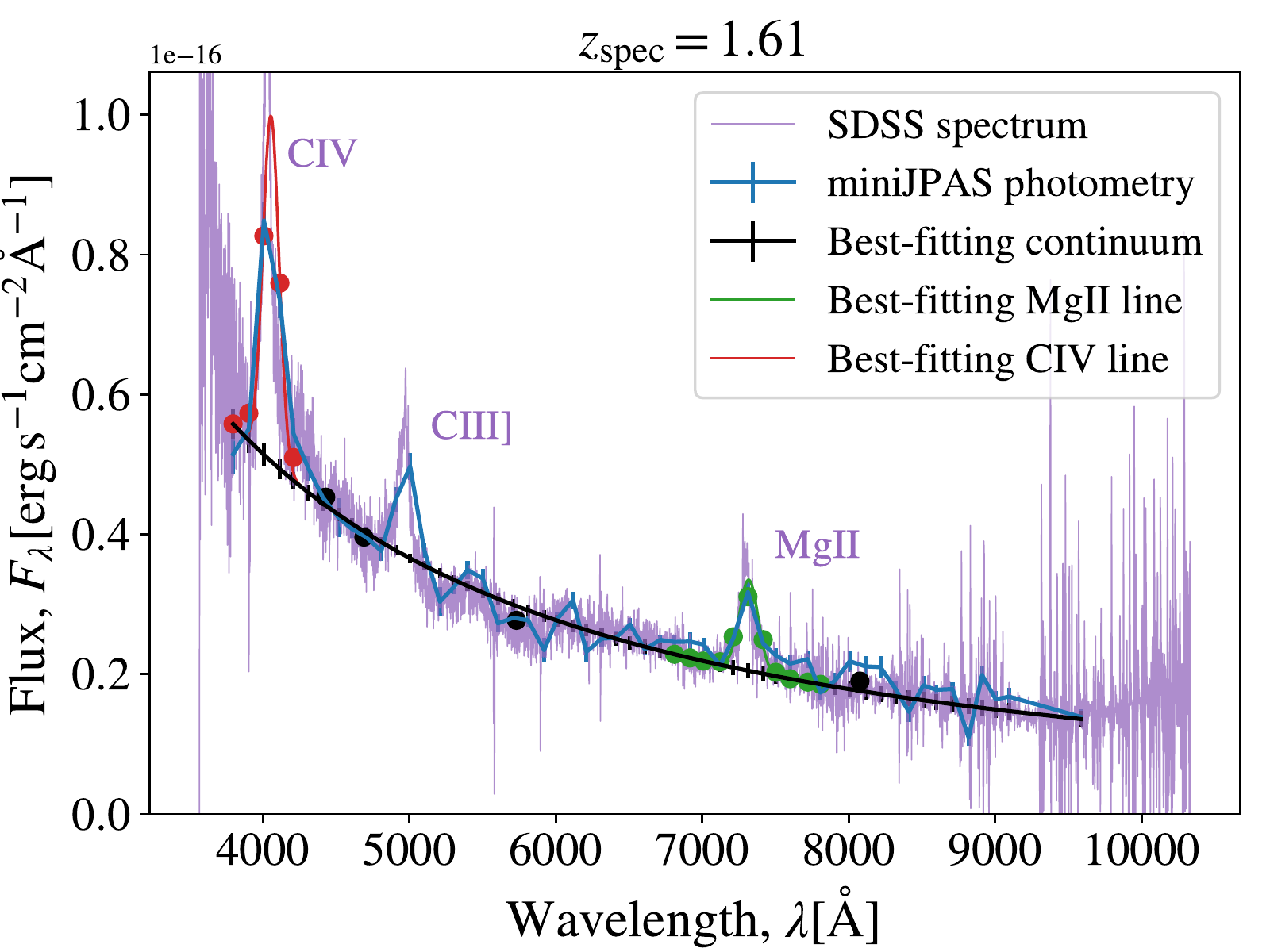}
    \includegraphics[width=\columnwidth]{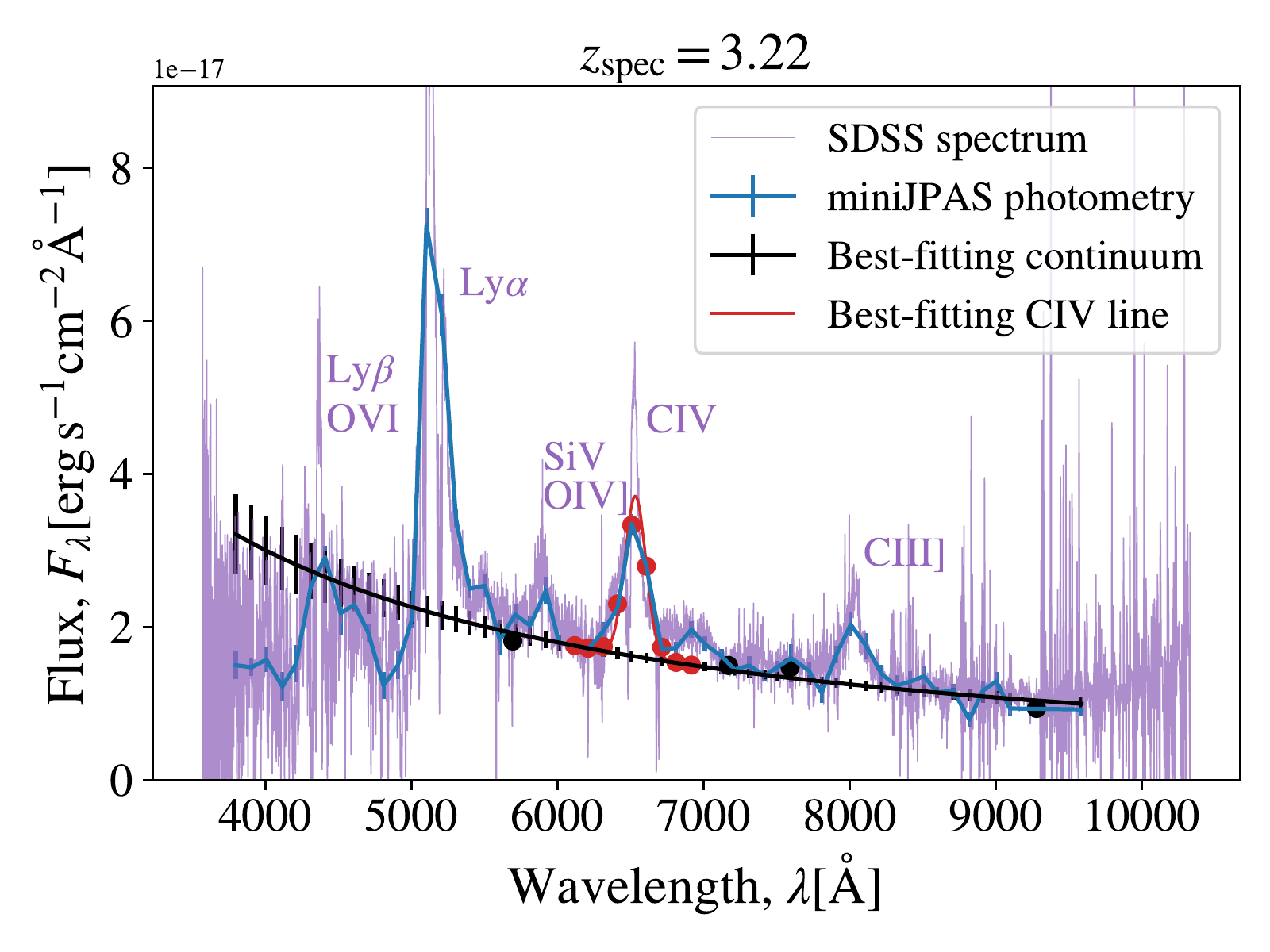}
    
    \caption{Photometric decomposition of miniJPAS data from three SDSS quasars. The top, middle, and bottom panels display the results for a quasar at low, intermediate, and high redshift, respectively. Blue and purple lines show miniJPAS photospectra and SDSS spectra, respectively, black lines display best-fitting continua, and orange, green, and red lines denote best-fitting \hbeta, \mgii, and \civ emission lines. Error bars show $1\sigma$-equivalent uncertainties. Despite the limited spectral resolution of J--PAS filters, we can readily see that best-fitting lines precisely capture the broad component of quasar emission lines in SDSS spectra.}
    \label{fig:example}
\end{figure}

\subsection{Continuum emission}
\label{sec:model_continuum}

The size of the BLR region presents a tight correlation with the luminosity of the quasar continuum emission, which is emitted by material in the accretion disk of the SMBH \citep{abramowicz2013_FoundationsBlackHole} and it is compatible with a power law from the optical to the near-UV \citep[e.g.][]{cristiani1990_CompositeSpectrumquasars, vandenberk01}. However, measuring the luminosity of the continuum is not straightforward; this is because the apparent continuum of a quasar results from the combination of the power-law continuum and other contributions such as unresolved emission and absorption lines, blended iron lines \citep[e.g.][]{veron-cetty2004_UnusualEmissionlinea}, Balmer continuum emission \citep[e.g.][]{wills1985_BroadEmissionfeatures}, host-galaxy contamination \citep[especially for faint sources; e.g.][]{shen2011_CatalogQuasarProperties, hernan-caballero2016_NeartomidInfraredspectrum}, dust reddening \citep[e.g.][]{hopkins2004_DustReddeningSloan}, and photometric errors. To alleviate the impact of these features on measurements of the monochromatic continuum luminosity, we first fit a power-law model to the apparent continuum emission, and then we measure the monochromatic luminosity from the best-fitting model.

To estimate the power-law continuum, we start by selecting a set of rest-frame wavelengths $\lambda_w$ not presenting strong emission features in their surroundings and sampling observer-frame photospectra in a sufficiently dense fashion up to $z=5$. We find that the wavelengths $\lambda_w=1350,$ 1700, 1800, 2200, 3100, 4000, 4200, and $5200\angstrom$ satisfy both criteria: these wavelengths present a separation of at least $100\angstrom$ from strong emission lines according to quasar composite spectra \citep[e.g.][]{vandenberk01} and at least two $\lambda_w$ fall within the J--PAS wavelength range up to $z\simeq5$. We continue by identifying the J--PAS narrow band with the closest pivot wavelength\footnote{As defined in Eq.~A11 of \citet{tokunaga2005_MaunaKeaObservatories}.} to the observer-frame value of each $\lambda_w$ for every source. Then, we compute the median flux of each selected band and those immediately preceding and succeeding; the resulting values approximate the continuum emission. We note that this approach is largely insensitive to redshift errors perturbing $\lambda_w$ less than half the width of survey filters (see Appendix~\ref{app:errors}).

Even though the selected $\lambda_w$ are not close to strong emission features, we find it necessary to apply some corrections to account for the impact of spectral features biasing high the continuum emission. By comparing SDSS spectra and miniJPAS photospectra, we find that the impact of the \ion{O}{iv}]-\ion{Si}{iv}~$\lambda\lambda1397.2,\,1402.8$ complex and blends of \ion{Fe}{ii} line emission redwards of the \mgii line is alleviated on average by reducing 10\% the flux at $\lambda_w=1350$ and $3100\angstrom$, respectively. Furthermore, we find that reducing 10\% the flux at $\lambda_w=5200\angstrom$ partially corrects for the change in the continuum slope starting at $\lambda\simeq5000\angstrom$, which is caused by a combination of host-galaxy contamination and emission from hot dust \citep[see][and references therein]{vandenberk01}. Both corrections enable a better estimation of the quasar continuum emission from the optical to the near-UV. On the other hand, we do not explicitly correct the quasar continuum for the impact of host-galaxy contamination because the quasars in the validation catalogue are brighter than $r=21$ and present redshift higher than $z=0.5$ (see Sect. \ref{sec:data_sample}), and the host-galaxy emission is increasingly weaker for brighter sources at higher redshift \citep[e.g.][]{shen2011_CatalogQuasarProperties}. We note that the results are weakly sensitive to all these corrections because we use multiple $\lambda_w$ to estimate the continuum emission for each source.

We use the publicly available affine invariant Markov chain Monte Carlo (MCMC) ensemble sampler {\sc emcee} \citep{foremanmackey13}\footnote{\url{https://emcee.readthedocs.io/en/stable/}} to compute the best-fitting power-law model to the continuum of each source, $f_\mathrm{cont} = f_0 \lambda^{\alpha_\lambda},$ where $f_0$ and $\alpha_\lambda$ are the normalisation and spectral index of the power law, respectively. This process works as follows. For each step of the Markov chain, {\sc emcee} draws a new value of the previous two parameters, convolves the resulting continuum with the J--PAS filter system, and compares the simulated and actual continuum emission to obtain the likelihood of the selected parameters. We ran the code using 100 independent chains of 150 steps, a burn-in phase of 75 steps, and broad uniform priors ($\alpha_\lambda\in[-3.5,\,3.5]$). We verify that this configuration results in a robust sampling of the posterior. To determine the best-fitting monochromatic continuum luminosity at a particular wavelength and its error, we first compute the luminosity of the continuum at such a wavelength from every accepted step of the MCMC chains. Then, we obtain the best-fitting solution and its uncertainty by computing the median and semi-amplitude of the range enclosing the 16 and 84th percentiles of the resulting values, respectively.

In top, middle, and bottom panels of Fig.~\ref{fig:example}, we display the photometric decomposition of miniJPAS data from three SDSS quasars at $z=0.68$, 1.61, and 3.22, respectively. The apparent magnitude of these sources is $r=18.1$, 20.1, and 20.3, their SDSS ID 7339--56722--108, 7339--56722--153, and 7339--56722--147, and their miniJPAS ID 00853, 15867, and 14873. Even though the spectral resolution of narrow-band filters is not high enough to resolve narrow spectral features, we can readily see that miniJPAS photospectra resolve broad emission lines precisely.

We find that the best-fitting continua follow the SDSS-observed continua closely, particularly for wavelengths not contaminated by important spectral features. Black dots indicate the values used to compute the best-fitting continua; as explained above, we select these wavelength intervals because the strongest quasar emission lines do not contaminate their flux. On the other hand, weaker spectral features affect some of these wavelength intervals. In the top panel, the $3000\angstrom$ bump \citep{grandi1982_3000Bumpquasars, oke1984_3000Bumpquasars, wills1985_BroadEmissionfeatures} and the change in the continuum slope near $\lambda=5000\angstrom$ \citep{vandenberk01} modify the flux of the black dots immediately redwards \mgii and \hbeta, respectively; nevertheless, we can readily see that the flux corrections mentioned above alleviate the impact of these features.


\subsection{Emission lines}
\label{sec:model_lines}


Among all quasar emission lines, we are primarily interested in \hbeta~$\lambda4861$, \mgii~$\lambda2798$, and \civ~$\lambda1549$ because these lines present EWs large enough to significantly modify narrow-band photometry and are calibrated to compute SMBH virial masses \citep[e.g.][]{kaspi2000_ReverberationMeasurements17, kaspi2005_RelationshipLuminosityBroadLine, vestergaard2006_DeterminingCentralBlack, shen2011_CatalogQuasarProperties, bentz2013_LowluminosityEndRadiusLuminosity}. In addition, we can detect at least one of these lines from the local universe up to $z=5$ using the J--PAS filter system, which enables a continuous estimation of SMBH masses up to such a redshift. In the following we describe our approach to extracting the properties of these emission lines.

For each source, we start by identifying the J--PAS bands with pivot wavelengths within the rest-frame intervals [4700, 5100]\angstrom, [2600, 3000]\angstrom, and [1450, 1630]\angstrom for the analysis of \hbeta, \mgii, and \civ, respectively. The widths of these intervals are $\Delta\lambda\simeq2.5$, 4.3, and $3.5\times10^4\,\kms,$ which are wide enough to encompass the broadest quasar emission lines almost entirely \citep[e.g.][]{rakshit2020_SpectralPropertiesquasars}. We restrict our analysis to emission lines with observer-frame central wavelength within the interval $[4000,\, 8900]\angstrom$ to ensure correct sampling of line wings. Then, we compute the relative difference between the miniJPAS photometry and best-fitting continuum emission for the selected bands (see Sect. \ref{sec:model_continuum}), producing a line-only spectrum. 

Spectral decomposition methods usually consider multiple broad and narrow components to recover the shape of emission line profiles more precisely \citep[e.g.][]{greene2005_EstimatingBlackHole, shen2011_CatalogQuasarProperties, guo2019_ConstrainingSubparsecbinary}. However, to avoid degeneracies between different components due to the limited spectral resolution of J--PAS photospectra, we use a single Gaussian to compute the amplitude, centre, and width of each emission line from the line-only spectrum. We do so using the MCMC sampler {\sc emcee} (see also Sect. \ref{sec:model_continuum}): for each step of the Markov chain, {\sc emcee} draws a new value for the amplitude, centre, and width of the target emission line, convolves the resulting line with the J--PAS filter system, and compares the simulated line and the line-only spectrum to obtain the likelihood of the selected parameters. For each line, we ran the code using 75 independent chains of 500 steps, a burn-in phase of 150 steps, and broad uniform priors. Specifically, we allow the line centre to move as much as an observer-frame distance of $75\angstrom$ from the rest-frame position of the target line, which corresponds to approximately half the width of a narrow-band J--PAS filter. This wide prior in the line centre aims to accommodate for possible velocity shifts or redshift errors. We compute the best-fitting value and error of line properties following the same strategy as for the continuum luminosity in Sect. \ref{sec:model_continuum}.

We find that the [\ion{O}{iii}]~$\lambda\lambda4958.9,\,5006.8$ complex and blended iron lines hinder the correct estimation of line properties for \hbeta and \mgii, respectively. Spectral methods usually model these features; however, the spectral resolution of J--PAS is too coarse to follow this approach. By comparing SDSS spectra and miniJPAS photospectra, we find that we can mitigate the overall impact of these features on line fits by reducing 50, 50, 25, and 50\% the flux of the J--PAS bands with pivot rest-frame wavelength closest to $\lambda=2700$, 2950, 4960, and $5008\angstrom$, respectively. The \hbeta correction is essential for all sources because the spectral resolution of J--PAS bands is not high enough to resolve \hbeta and the [\ion{O}{iii}]~$\lambda\lambda4958.9,\,5006.8$ complex separately. The \mgii correction only improves the results for lines broader than $\sim 8000\,\kms$ because for narrower lines the spectral resolution of J--PAS is high enough to resolve \mgii and blended iron lines separately. Not introducing these corrections results in overestimating the width of emission lines.

In Fig.~\ref{fig:example}, we show the best-fitting emission line models to the broad lines of example quasars. By comparing the result of photometric measurements and SDSS spectra, we can readily see that best-fitting lines capture the broad component of \hbeta, \mgii, and \civ precisely. This level of agreement is remarkable given the significant difference in spectral resolution between SDSS spectra and J--PAS photospectra, with average spectral resolutions of $R\simeq1800$ and 60 \cite[respectively;][]{york00, benitez2014_JPASJavalambrePhysicsAccelerated}. The spectral resolution of J--PAS photospectra suggests that we can only resolve lines broader than $\sim 5000\,\kms$; however, this calculation does not account for the overlapping of the transmission curve of adjacent J--PAS filters.


\subsection{SMBH virial masses}
\label{sec:model_masses}

At cosmological distances, the standard approach to compute SMBH masses assumes that the BLR is virialised and that the SMBH gravitational field dominates the motion of gas clouds in this region. Single-epoch spectroscopy computes virial masses by leveraging the tight correlation between continuum luminosity and BLR size \citep[e.g.][]{kaspi2000_ReverberationMeasurements17, bentz2009_RadiusLuminosityRelationshipActive},
\begin{equation}
    \label{eq:mass_estimator}
    \log\left(\frac{M_\mathrm{BH}}{\Msun}\right) = A + B \log\left(\frac{\lambda L_\lambda}{10^{44}\ergs}\right) + 2\log\left(\frac{\mathrm{FWHM}}{\kms}\right),
\end{equation}
where FWHM stands for the full width at half maximum of broad emission lines, $\lambda L_\lambda$ refers to the monochromatic continuum luminosity -- typically measured over a spectral region adjacent to the respective broad emission line --, and $A$ and $B$ are virial coefficients calibrated using either sources with both SES and RM measurements \citep[e.g.][]{kaspi2000_ReverberationMeasurements17, bentz2013_LowluminosityEndRadiusLuminosity} or internally based on the availability of multiple lines for the same source \citep[e.g.][]{mclure2002_MeasuringBlackhole, vestergaard2006_DeterminingCentralBlack,trakhtenbrot2012_BlackHolegrowtha, marinello2020_VLTSINFONIstudy}. We compute \hbeta-, \mgii-, and \civ-based virial masses using the continuum luminosity at $\lambda=5100$, 3000, and $1350\angstrom$, respectively, and the same virial coefficients as SDSS-based quasar catalogues \citep{shen2011_CatalogQuasarProperties, rakshit2020_SpectralPropertiesquasars}: $A=0.91$ and $B=0.50$ for \hbeta \citep{vestergaard2006_DeterminingCentralBlack}, $A=0.74$ and $B=0.62$ for \mgii \citep{shen2011_CatalogQuasarProperties, trakhtenbrot2012_BlackHolegrowtha}, and $A=0.66$ and $B=0.53$ for \civ \citep{vestergaard2006_DeterminingCentralBlack}. Consequently, SMBH masses depend approximately four times more strongly on the FWHM than continuum luminosity measurements.

Different calibrations or spectral decomposition techniques may result in differences as large as 0.4 dex between RM and SES masses \citep[e.g.][]{collin2006_SystematicEffectsmeasurement, kelly2009_AreVariationsQuasar, shen2013_MassQuasars, bonta2020_SloanDigitalSky}; motivated by this, some works recalibrate virial coefficients using RM and SES measurements from the same spectral decomposition code to reduce these errors. It is also worth noting that the precision of SMBH mass estimates depends on the emission line used during the inference process: \hbeta-based masses present a scatter of 0.3 and 0.5 dex relative to \mgii- and \civ-based masses, respectively \citep{trakhtenbrot2012_BlackHolegrowtha}.

\begin{figure*}
    \centering
    \includegraphics[width=0.95\columnwidth]{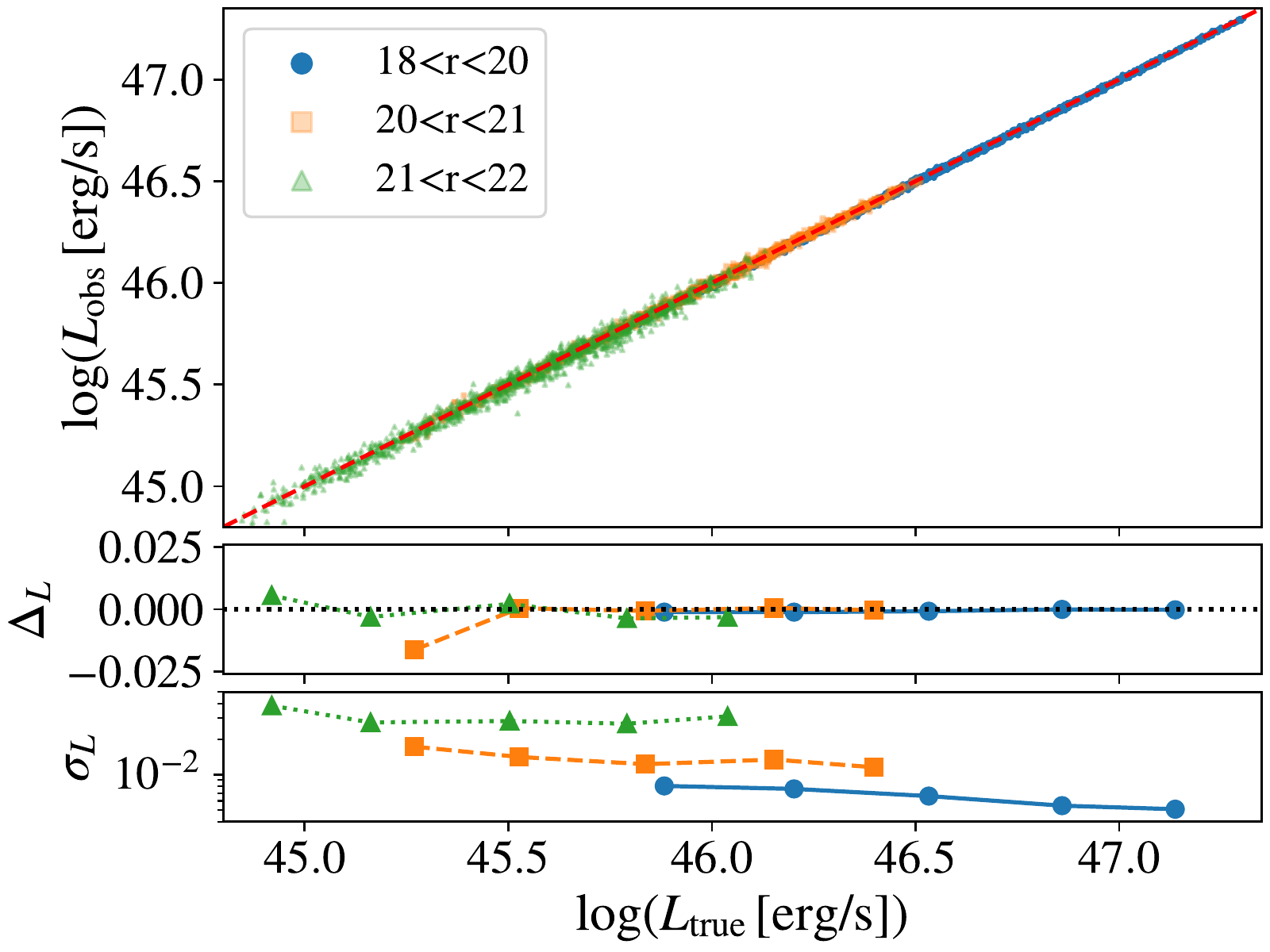}
    \includegraphics[width=0.95\columnwidth]{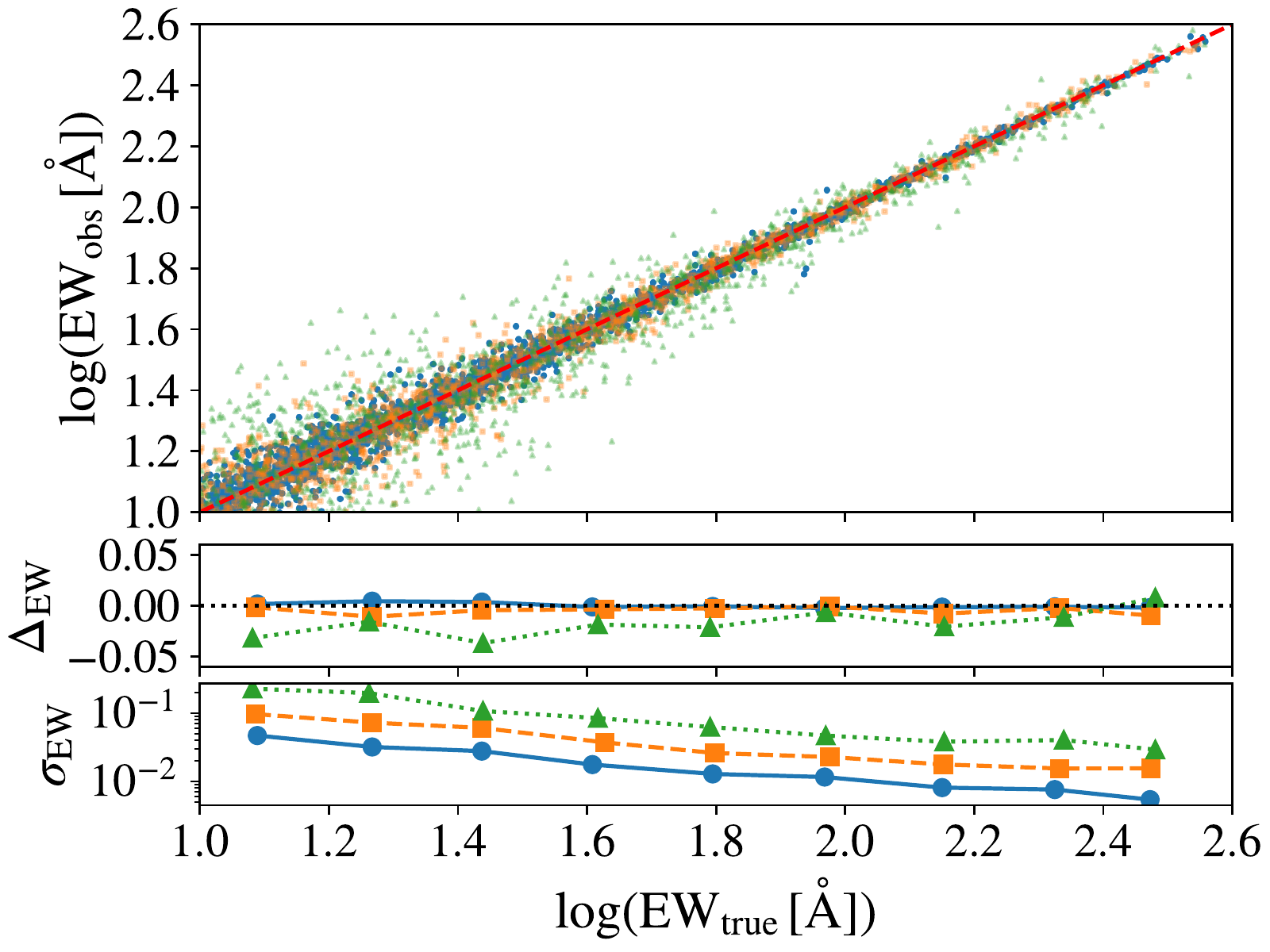}
    
    \includegraphics[width=0.95\columnwidth]{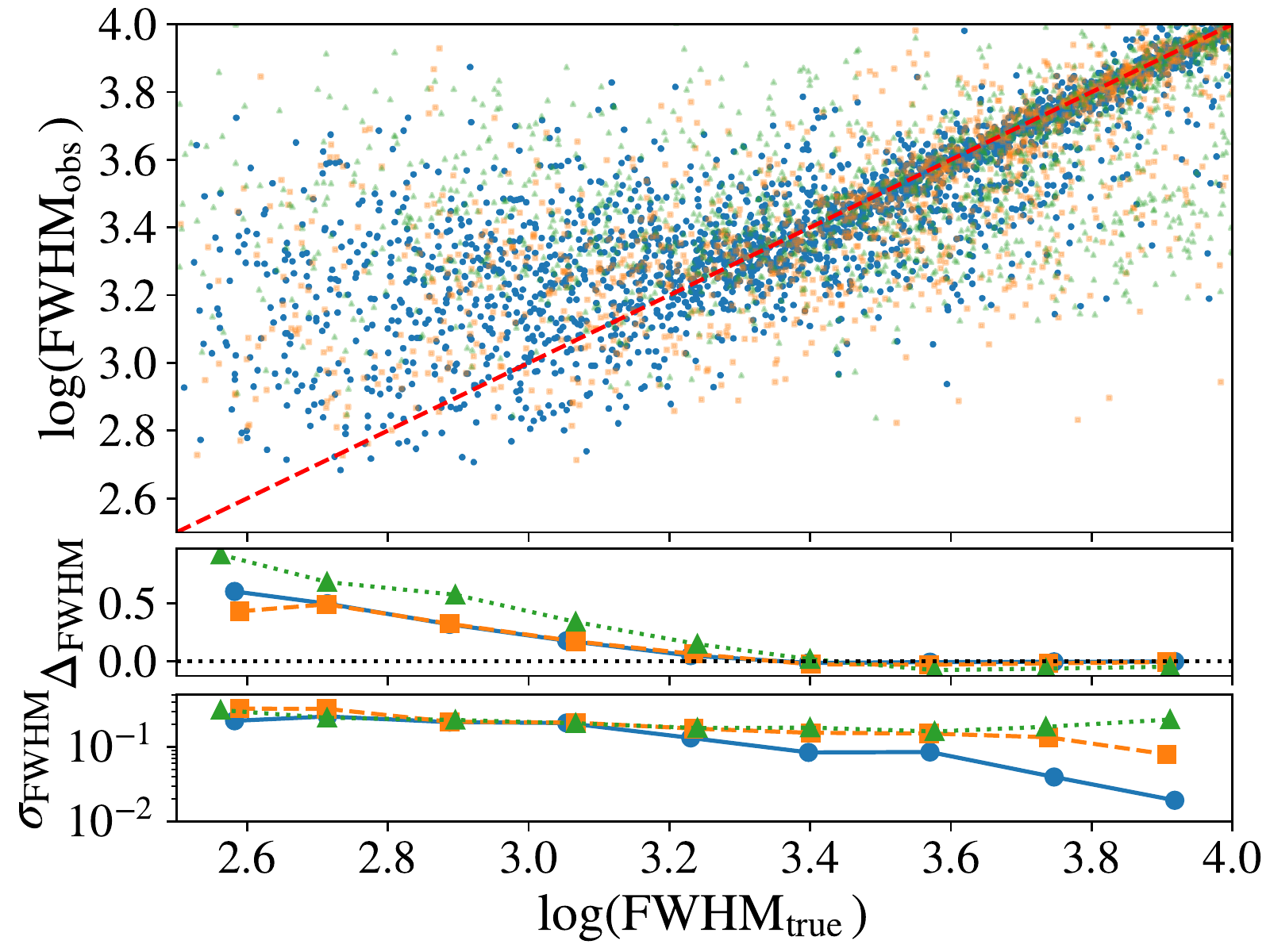}
    \includegraphics[width=0.95\columnwidth]{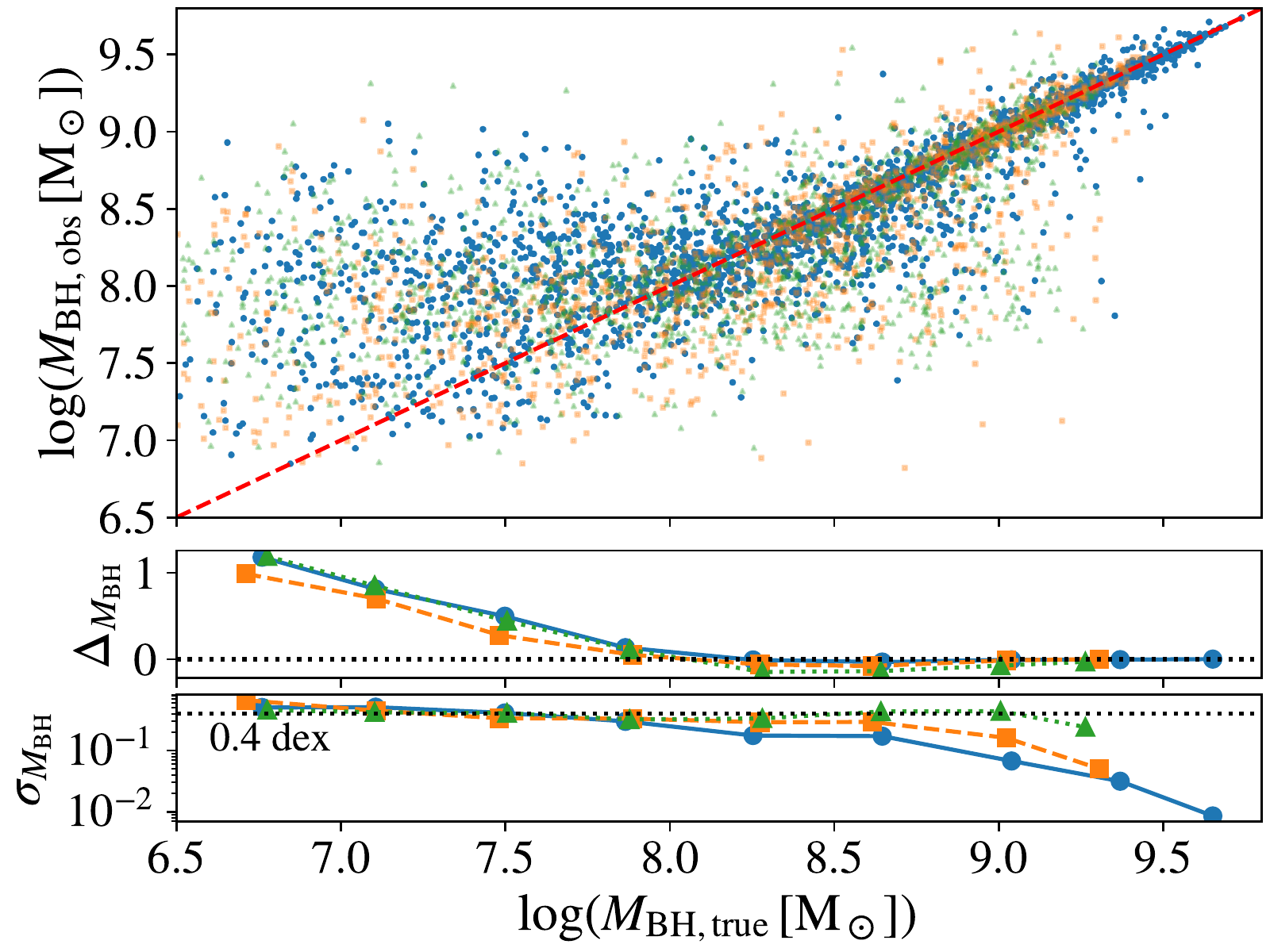}
    \caption{SEP measurements of $\lambda L_\lambda(1350\angstrom)$ continuum luminosity (top-left panel), \civ EW (top-right panel), \civ FWHM (bottom-left panel), and \civ-based SMBH mass (bottom-right panel) from simulated J--PAS sources. Blue dots, orange squares, and green triangles show the results for sources with $18<r<20$, $20<r<21$, and $21<r<22$, respectively, and dashed red lines indicate a one-to-one relation between actual and measured properties. The middle and bottom sub-panels display the mean and standard deviation of the logarithmic difference between actual and measured properties. We find that our method yields unbiased FWHM measurements only for lines broader than $\simeq1500\,\kms$ due to the limited spectral resolution of J--PAS photospectra, which prevents us from measuring the mass of sources with $\log(M_\mathrm{BH}/\Msun)\lesssim8$ in an unbiased fashion. We find similar results for \hbeta and \mgii.}
    \label{fig:forecast}
\end{figure*}

To compute SEP masses, we use continuum luminosities and line widths estimated from J--PAS photospectra (see Sects. \ref{sec:model_continuum} and \ref{sec:model_lines}, respectively). Ideally, we would recalibrate Eq.~\ref{eq:mass_estimator} coefficients using sources presenting both RM and SEP mass measurements; however, miniJPAS did not observe any quasar with RM measurements. In Sect. \ref{sec:results}, we resort to recalibrating these coefficients using sources with both single-epoch spectroscopic and photometric measurements; to do so, we use the gradient-based BFGS algorithm \citep{broyden1970_ConvergenceClassDoublerank, fletcher1970_NewApproachvariable, goldfarb1970_FamilyVariablemetricmethods, shanno1970_OptimalConditioningquasiNewton} implemented in {\sc scipy} \citep{virtanen2020_SciPyFundamentalalgorithms}. This recalibration has two important benefits: it absorbs systematic differences between spectroscopic and photometric measurements of both continuum luminosity, which may be caused by over- or under-estimating the correction from aperture to total magnitudes (see Sect. \ref{sec:data_sample}), and line widths, which may appear due to the different number of components used to fit line profiles (see Sect. \ref{sec:model_lines}).


\subsection{Tests using idealised simulations}
\label{sec:model_mock}

In order to thoroughly test our methodology, we proceed to study the feasibility of using SEP to measure SMBH masses using simulated observations of J--PAS quasars.

Modelling the spectral energy distribution of a quasar is challenging due to correlations between multiple quasar properties \citep[e.g.][]{baldwin77, dong2009_EddingtonRatioGoverns, shen2011_CatalogQuasarProperties}, the presence of a panoply of weak and blended emission lines \citep[e.g.][]{davidson1979_EmissionLinesquasars}, the complex profile of some broad emission lines \citep[e.g.][]{nagao2006_EvolutionBroadlineregion, shen2011_CatalogQuasarProperties, kollatschny2013_ShapeBroadlineprofilesa}, the diversity of quasar continua \citep[e.g.][]{jensen16}, emission line velocity shifts \citep[e.g.][]{richards2011_UnificationLuminousType, shen2016_SloanDigitalSky}, and host-galaxy contamination \citep[e.g.][]{vandenberk01}. Instead of accounting for all these effects, we proceed to generate idealised quasar observations by modelling the quasar continuum emission as a power law and each broad emission line as a single Gaussian. This approach ensures that the only sources of uncertainty affecting SEP masses are the limited resolution of the J--PAS filter system and the level of photometric errors expected for this survey. In Sect. \ref{sec:results}, we study the aggregated impact of all the aforementioned effects and others such as variability by comparing SES and SEP measurements from quasars in the miniJPAS validation sample (Sect. \ref{sec:data_sample}).

For each source, we first modelled the continuum emission using a power law of index $\alpha_\lambda=-1.56$, which provides an excellent fit to the quasar continuum emission from Ly$\alpha$ to \hbeta \citep{vandenberk01}. Then, we added the emission lines \hbeta, \mgii, and \civ to the continuum emission, which we modelled using a single Gaussian function with the same width. After that, we redshifted the resulting spectral energy distribution from rest- to observer-frame, and we convolved it with the J--PAS filter system. Finally, we perturbed mock photospectra according to the level of photometric uncertainties expected for the J--PAS survey \citep{benitez2014_JPASJavalambrePhysicsAccelerated}. In summary, four free parameters characterise each simulated photospectra: continuum $r$-band magnitude, rest-frame EW, FWHM, and redshift. To generate mock observations, we first drew 10\,500 random combinations of these parameters within the intervals $z\in[0.01,\,4.50]$, $r\in[18,\,23]$, $\log(\mathrm{EW/\angstrom})\in[1,\,2.6]$, and  $\log(\mathrm{FWHM}/\kms)\in[2.5,\,4.0]$ using Latin hypercube sampling \citep{mckay1979_ComparisonThreemethods}. Then, we followed the previously mentioned strategy to generate a J--PAS photospectra for each combination of parameters.

In Fig.~\ref{fig:forecast}, we compare the SEP measurements from mock photospectra and the input quantities used to simulate these. We note that we analysed mock observations using a slightly modified version of the model described in Sect. \ref{sec:model}: we did not apply any correction to either the continuum emission or emission lines because simulated photospectra do not incorporate any contaminant. Blue dots, orange squares, and green triangles indicate the results for sources with $18<r<20$, $20<r<21$, and $21<r<22$, respectively, and red dashed lines indicate the 1:1~relations between input and measured quantities. In each panel, the middle and bottom sub-panels display the mean and standard deviation of the logarithmic difference between input and measured values. As expected, the precision of all measurements is increasingly larger for brighter sources as the relative impact of the associated photometric errors decreases.

In the top-left and top-right panels, we display measurements of the monochromatic continuum luminosity at $1350\angstrom$ and the EW of \civ, respectively. As we can see, the precision of the measurements increases for more luminous sources and stronger lines; this is because the relative impact of photometric errors decreases with both the brightness of the continuum and the strength of emission lines. We also find that our model only delivers unbiased measurements for sources brighter than $r\simeq21$; motivated by this, we only consider sources brighter than such a magnitude to estimate the performance of SEP in Sect. \ref{sec:results}.

In the bottom-left panel, we show measurements of the FWHM of \civ. We find that the precision of these measurements increases for broader lines; this is because the J--PAS filter system resolves wider lines with a larger number of bands. The most prominent feature of this panel is that FWHM measurements display a systematic bias for emission lines narrower than ${\approx}1500\,\kms$. We can understand this trend in terms of the limited spectral resolution of the J--PAS filter system: on average, emission lines narrower than ${\approx}1500\,\kms$ perturb the flux of a single J--PAS band, which causes a complete degeneracy between the amplitude and FWHM of the best-fitting line. For such lines, our method delivers practically any result between the actual width and the minimum one resolvable, thereby biasing high FWHM measurements. In Appendix~\ref{app:back}, we show that backward-modelling observations result in systematic biases even for lines as wide as ${\approx}10\,000\,\kms$; therefore, forward-modelling quasar observations enables us to push this limit by almost an order of magnitude. We note that our naive estimate for the width of the narrowest line resolvable by J--PAS is approximately three times larger than the actual value (see Sect. \ref{sec:model_lines}).

In the bottom-right panel, we display measurements of \civ-based SMBH mass. We note that we compute SMBH masses using the virial coefficients quoted in Sect. \ref{sec:model_masses}. As we can see, our method yields unbiased masses only for sources with $\log(M_\mathrm{BH}/\Msun)\gtrsim8$, reflecting the biased FWHM measurements for lines narrower than $\simeq1500\,\kms$. For sources with FWHM larger than the aforementioned threshold, we find that SEP masses show no systematic bias and that the precision of these ranges from 0.4 to 0.01 dex for SMBH with masses from $\log(M_\mathrm{BH}/\Msun)\simeq8$ to 9.75, respectively. We also find that the precision of these measurements increases with the SMBH mass for sources brighter than $r=21$, which is explained by the increasing precision of both continuum luminosity and FWHM measurements for sources with brighter continua and broader lines. We note that we find similar results for \hbeta- and \mgii-based measurements.

\begin{figure}
    \centering
    \includegraphics[width=\columnwidth]{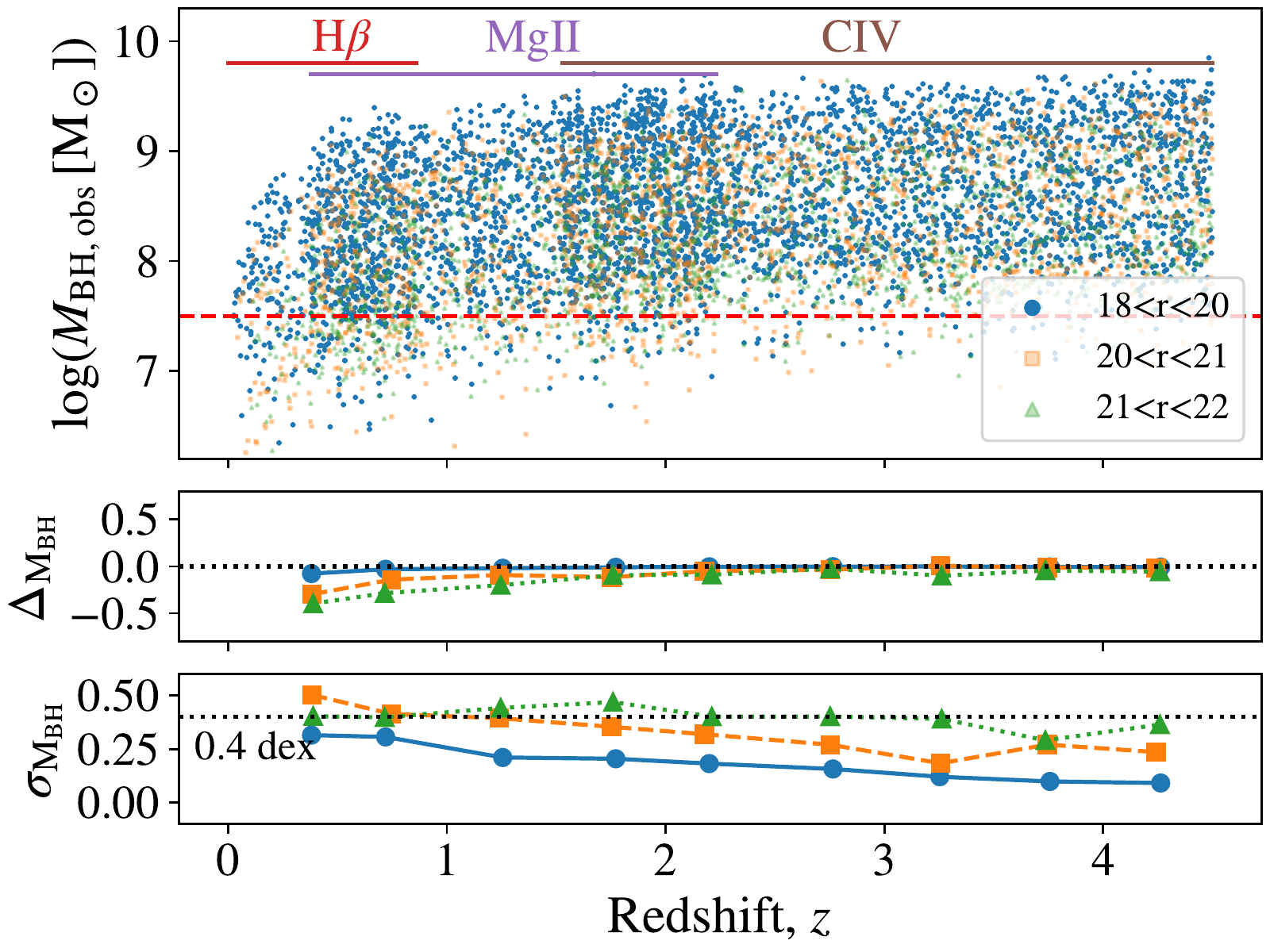}
    \caption{SEP masses from simulated J--PAS observations of sources with actual SMBH masses larger than $\log(M_\mathrm{BH}/\Msun)=7.5$. The dashed red line indicates this threshold. As we can see, the combination of \hbeta, \mgii, and \civ enables SMBH masses from $z=0$ to 5 to be continuously measured.}
    \label{fig:forecast_bhmz}
\end{figure}

In Fig.~\ref{fig:forecast_bhmz}, we display SEP mass measurements from \hbeta, \mgii, and \civ as a function of redshift. As we can see, these lines enable SMBH masses from $z=0$ to 5 to be measured continuously. We only display results for sources with an actual mass larger than $\log(M_\mathrm{BH}/\Msun)=7.5$; we set this threshold to understand better the impact of the limited resolution of the J--PAS filter system on measurements for low mass sources. We find that our model yields unbiased SMBH masses for sources with both $r<20$ and $\log(M_\mathrm{BH}/\Msun)>7.5$ across the whole redshift range. On the other hand, sources with $r>20$ present slightly biased masses below $z=1,$ especially those with less massive SMBHs. This is due to the combination of two effects:  the precision of SEP measurements increases with both apparent magnitude and line width, and at fixed FWHM (in $\kms$), the observer-frame line width (in\,$\angstrom$) grows with redshift as $1+z$. We check that our model yields unbiased SMBH mass estimates for sources with emission lines broader than an observer-frame width of $\mathrm{FWHM}=30\angstrom$.


\section{Results}
\label{sec:results}


In Sect. \ref{sec:model}, we describe SEP and we validate our methodology using simulated J--PAS photospectra. In this section we characterise the precision of this technique by comparing SEP and SES measurements for 54 SDSS quasars observed by the miniJPAS survey (see Sect. \ref{sec:data_sample}). In Sects. \ref{sec:results_continuum}, \ref{sec:results_lines}, and \ref{sec:results_masses}, we show the results for continuum luminosities, emission line properties, and SMBH virial masses, respectively, and we gather these in Table~\ref{tab:data}.


\begin{table*}
    \caption{SEP measurements for some sources in the miniJPAS validation catalogue. An extended version of this table containing SEP and SES measurements for all sources in the validation catalogue can be found \href{https://jchavesmontero.github.io/jchavesmontero/projects/agn/minijpas.fits}{here}.}
    \label{tab:data}
    \begin{center}
    \begin{tabular}{cccccccccc}
    \hline
    $z$ & \multicolumn{3}{c}{$\log(L_\lambda[\mathrm{erg/s}])$} & \multicolumn{3}{c}{$\log(\mathrm{FWHM}[\kms])$} & \multicolumn{3}{c}{$\log(M_\mathrm{BH}[\Msun])$}\\
    & $5100\angstrom$ & $3000\angstrom$ & $1350\angstrom$ & \hbeta & \mgii & \civ & \hbeta & \mgii & \civ\\
    \hline
    0.482 & $44.16^{+0.03}_{-0.03}$ & $44.23^{+0.02}_{-0.02}$ & -- & $3.81^{+0.29}_{-0.37}$ & $3.61^{+0.52}_{-0.55}$ & -- & $8.53\pm0.66$ & $7.65\pm1.06$ & -- \\
    0.531 & $44.10^{+0.04}_{-0.04}$ & $43.86^{+0.04}_{-0.04}$ & -- & $3.91^{+0.15}_{-0.36}$ & $3.87^{+0.18}_{-0.16}$ & -- & $8.73\pm0.51$ & $7.94\pm0.34$ & -- \\
    0.548 & $44.32^{+0.03}_{-0.03}$ & $44.29^{+0.04}_{-0.04}$ & -- & $3.87^{+0.18}_{-0.38}$ & $3.74^{+0.24}_{-0.24}$ & -- & $8.71\pm0.56$ & $7.94\pm0.49$ & -- \\
    0.600 & $44.32^{+0.05}_{-0.05}$ & $43.74^{+0.08}_{-0.07}$ & -- & $3.60^{+0.47}_{-0.75}$ & $4.08^{+0.23}_{-0.25}$ & -- & $8.16\pm1.22$ & $8.30\pm0.47$ & -- \\
    0.602 & $44.39^{+0.02}_{-0.02}$ & $44.45^{+0.02}_{-0.02}$ & -- & $3.91^{+0.13}_{-0.16}$ & $4.03^{+0.11}_{-0.13}$ & -- & $8.81\pm0.29$ & $8.62\pm0.24$ & -- \\
    0.647 & $44.78^{+0.02}_{-0.02}$ & $44.89^{+0.02}_{-0.02}$ & -- & $3.80^{+0.18}_{-0.28}$ & $3.83^{+0.26}_{-0.52}$ & -- & $8.67\pm0.46$ & $8.47\pm0.78$ & -- \\
    0.676 & $45.21^{+0.02}_{-0.02}$ & $45.48^{+0.02}_{-0.02}$ & -- & $3.72^{+0.10}_{-0.17}$ & $3.86^{+0.24}_{-0.66}$ & -- & $8.62\pm0.28$ & $8.89\pm0.90$ & -- \\
    0.676 & $44.47^{+0.04}_{-0.04}$ & $44.45^{+0.05}_{-0.05}$ & -- & $3.65^{+0.38}_{-0.73}$ & $3.85^{+0.47}_{-0.82}$ & -- & $8.31\pm1.11$ & $8.26\pm1.29$ & -- \\
    0.719 & $44.51^{+0.04}_{-0.04}$ & $44.61^{+0.03}_{-0.03}$ & -- & $3.83^{+0.26}_{-0.84}$ & $3.59^{+0.61}_{-0.58}$ & -- & $8.66\pm1.10$ & $7.84\pm1.18$ & -- \\
    0.808 & $44.46^{+0.04}_{-0.04}$ & $44.36^{+0.03}_{-0.03}$ & -- & $3.48^{+0.50}_{-0.68}$ & $4.09^{+0.21}_{-0.49}$ & -- & $7.96\pm1.18$ & $8.67\pm0.70$ & -- \\
    0.825 & $44.38^{+0.05}_{-0.05}$ & $44.38^{+0.04}_{-0.04}$ & -- & $3.55^{+0.42}_{-0.39}$ & $3.54^{+0.24}_{-0.18}$ & -- & $8.07\pm0.81$ & $7.60\pm0.42$ & -- \\
    0.884 & -- & $44.52^{+0.02}_{-0.02}$ & -- & -- & $3.87^{+0.14}_{-0.23}$ & -- & -- & $8.34\pm0.37$ & -- \\
    0.897 & -- & $44.68^{+0.03}_{-0.03}$ & -- & -- & $4.09^{+0.16}_{-0.25}$ & -- & -- & $8.88\pm0.41$ & -- \\
    0.986 & -- & $44.49^{+0.04}_{-0.04}$ & -- & -- & $4.05^{+0.14}_{-0.18}$ & -- & -- & $8.67\pm0.32$ & -- \\
    0.986 & -- & $44.75^{+0.02}_{-0.02}$ & -- & -- & $3.71^{+0.19}_{-0.28}$ & -- & -- & $8.16\pm0.46$ & -- \\
    1.002 & -- & $44.71^{+0.03}_{-0.03}$ & -- & -- & $3.89^{+0.33}_{-0.51}$ & -- & -- & $8.50\pm0.84$ & -- \\
    1.086 & -- & $44.65^{+0.03}_{-0.02}$ & -- & -- & $3.99^{+0.17}_{-0.28}$ & -- & -- & $8.66\pm0.45$ & -- \\
    1.194 & -- & $45.29^{+0.02}_{-0.01}$ & -- & -- & $4.01^{+0.09}_{-0.10}$ & -- & -- & $9.08\pm0.19$ & -- \\
    1.213 & -- & $45.64^{+0.01}_{-0.01}$ & -- & -- & $3.94^{+0.13}_{-0.21}$ & -- & -- & $9.14\pm0.35$ & -- \\
    1.223 & -- & $44.73^{+0.05}_{-0.04}$ & -- & -- & $4.13^{+0.12}_{-0.13}$ & -- & -- & $8.98\pm0.25$ & -- \\
    1.269 & -- & $45.20^{+0.02}_{-0.02}$ & -- & -- & $3.76^{+0.24}_{-0.46}$ & -- & -- & $8.52\pm0.70$ & -- \\
    1.286 & -- & $46.03^{+0.01}_{-0.01}$ & -- & -- & $4.04^{+0.05}_{-0.06}$ & -- & -- & $9.57\pm0.11$ & -- \\
    1.391 & -- & $45.31^{+0.02}_{-0.03}$ & -- & -- & $3.78^{+0.22}_{-0.31}$ & -- & -- & $8.62\pm0.53$ & -- \\
    1.394 & -- & $45.61^{+0.02}_{-0.02}$ & -- & -- & $3.49^{+0.31}_{-0.39}$ & -- & -- & $8.22\pm0.70$ & -- \\
    1.492 & -- & $45.44^{+0.03}_{-0.03}$ & -- & -- & $4.09^{+0.12}_{-0.16}$ & -- & -- & $9.33\pm0.29$ & -- \\
    1.514 & -- & $45.86^{+0.02}_{-0.02}$ & -- & -- & $3.99^{+0.11}_{-0.13}$ & -- & -- & $9.37\pm0.24$ & -- \\
    1.515 & -- & $45.07^{+0.05}_{-0.04}$ & -- & -- & $4.05^{+0.15}_{-0.15}$ & -- & -- & $9.01\pm0.29$ & -- \\
    1.584 & -- & $45.65^{+0.03}_{-0.03}$ & $45.96^{+0.02}_{-0.02}$ & -- & $4.04^{+0.09}_{-0.09}$ & $3.57^{+0.27}_{-0.27}$ & -- & $9.33\pm0.19$ & $8.55\pm0.54$ \\
    1.605 & -- & $45.43^{+0.03}_{-0.03}$ & $45.62^{+0.03}_{-0.03}$ & -- & $3.83^{+0.18}_{-0.22}$ & $3.64^{+0.19}_{-0.20}$ & -- & $8.80\pm0.41$ & $8.60\pm0.39$ \\
    1.647 & -- & $45.54^{+0.03}_{-0.03}$ & $45.86^{+0.03}_{-0.03}$ & -- & $4.00^{+0.11}_{-0.11}$ & $3.70^{+0.28}_{-0.33}$ & -- & $9.20\pm0.22$ & $8.79\pm0.61$ \\
    1.674 & -- & $45.05^{+0.12}_{-0.12}$ & $45.23^{+0.10}_{-0.10}$ & -- & $4.06^{+0.15}_{-0.40}$ & $3.97^{+0.27}_{-0.42}$ & -- & $9.03\pm0.55$ & $9.16\pm0.69$ \\
    1.685 & -- & $45.41^{+0.04}_{-0.04}$ & $45.51^{+0.03}_{-0.03}$ & -- & $3.89^{+0.12}_{-0.16}$ & $4.18^{+0.05}_{-0.06}$ & -- & $8.91\pm0.28$ & $9.64\pm0.11$ \\
    1.728 & -- & $45.15^{+0.08}_{-0.08}$ & $45.49^{+0.05}_{-0.05}$ & -- & $3.92^{+0.25}_{-0.37}$ & $3.63^{+0.34}_{-0.35}$ & -- & $8.80\pm0.62$ & $8.56\pm0.69$ \\
    1.743 & -- & $45.58^{+0.03}_{-0.03}$ & $45.79^{+0.03}_{-0.03}$ & -- & $4.10^{+0.11}_{-0.15}$ & $3.82^{+0.22}_{-0.37}$ & -- & $9.43\pm0.26$ & $9.00\pm0.59$ \\
    1.862 & -- & $45.60^{+0.05}_{-0.05}$ & $45.39^{+0.05}_{-0.05}$ & -- & $3.72^{+0.41}_{-0.80}$ & $4.17^{+0.24}_{-0.71}$ & -- & $8.68\pm1.21$ & $9.60\pm0.95$ \\
    1.902 & -- & $45.41^{+0.05}_{-0.05}$ & $45.65^{+0.02}_{-0.02}$ & -- & $3.51^{+0.45}_{-0.39}$ & $3.79^{+0.09}_{-0.08}$ & -- & $8.15\pm0.84$ & $8.90\pm0.16$ \\
    1.902 & -- & $45.74^{+0.03}_{-0.03}$ & $45.84^{+0.02}_{-0.02}$ & -- & $3.84^{+0.22}_{-0.57}$ & $3.64^{+0.22}_{-0.29}$ & -- & $9.00\pm0.78$ & $8.65\pm0.51$ \\
    1.960 & -- & $45.58^{+0.03}_{-0.03}$ & $45.73^{+0.02}_{-0.02}$ & -- & $3.78^{+0.24}_{-0.53}$ & $4.05^{+0.06}_{-0.07}$ & -- & $8.78\pm0.77$ & $9.45\pm0.13$ \\
    1.963 & -- & $45.38^{+0.04}_{-0.04}$ & $45.62^{+0.02}_{-0.02}$ & -- & $3.98^{+0.13}_{-0.16}$ & $3.73^{+0.19}_{-0.24}$ & -- & $9.07\pm0.29$ & $8.78\pm0.43$ \\
    2.003 & -- & $45.52^{+0.04}_{-0.04}$ & $45.63^{+0.03}_{-0.03}$ & -- & $3.94^{+0.11}_{-0.13}$ & $3.95^{+0.14}_{-0.25}$ & -- & $9.06\pm0.24$ & $9.23\pm0.39$ \\
    2.031 & -- & $45.66^{+0.02}_{-0.03}$ & $45.60^{+0.02}_{-0.02}$ & -- & $4.11^{+0.08}_{-0.12}$ & $4.01^{+0.07}_{-0.06}$ & -- & $9.49\pm0.20$ & $9.34\pm0.13$ \\
    2.033 & -- & $45.45^{+0.04}_{-0.04}$ & $45.70^{+0.04}_{-0.03}$ & -- & $4.07^{+0.09}_{-0.10}$ & $3.75^{+0.16}_{-0.18}$ & -- & $9.28\pm0.19$ & $8.83\pm0.34$ \\
    2.041 & -- & $45.54^{+0.03}_{-0.03}$ & $45.70^{+0.03}_{-0.03}$ & -- & $3.86^{+0.24}_{-0.32}$ & $3.59^{+0.48}_{-0.52}$ & -- & $8.93\pm0.56$ & $8.52\pm1.00$ \\
    2.305 & -- & -- & $45.81^{+0.05}_{-0.05}$ & -- & -- & $3.95^{+0.11}_{-0.12}$ & -- & -- & $9.26\pm0.23$ \\
    2.306 & -- & -- & $46.17^{+0.02}_{-0.02}$ & -- & -- & $4.07^{+0.08}_{-0.09}$ & -- & -- & $9.61\pm0.17$ \\
    2.351 & -- & -- & $45.58^{+0.07}_{-0.07}$ & -- & -- & $4.08^{+0.16}_{-0.22}$ & -- & -- & $9.48\pm0.37$ \\
    2.463 & -- & -- & $46.63^{+0.02}_{-0.02}$ & -- & -- & $3.95^{+0.07}_{-0.08}$ & -- & -- & $9.49\pm0.15$ \\
    2.581 & -- & -- & $46.10^{+0.02}_{-0.02}$ & -- & -- & $3.85^{+0.12}_{-0.29}$ & -- & -- & $9.15\pm0.41$ \\
    2.591 & -- & -- & $46.02^{+0.02}_{-0.02}$ & -- & -- & $4.11^{+0.06}_{-0.07}$ & -- & -- & $9.64\pm0.14$ \\
    2.594 & -- & -- & $45.99^{+0.02}_{-0.02}$ & -- & -- & $3.96^{+0.09}_{-0.11}$ & -- & -- & $9.34\pm0.20$ \\
    \hline
    \end{tabular}
    \end{center}
\end{table*}

\begin{figure}
    \centering
    \includegraphics[width=\columnwidth]{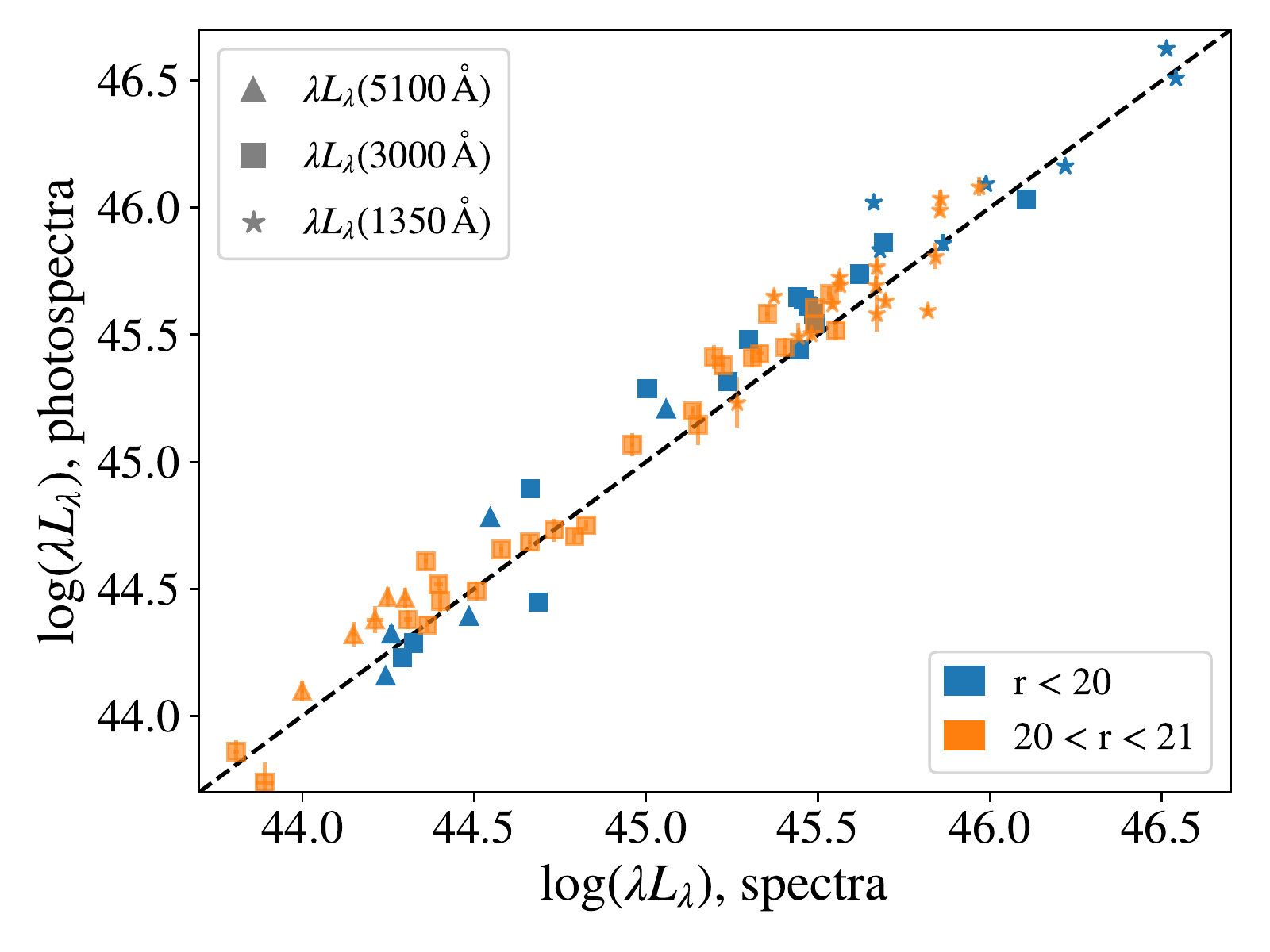}
    \caption{Continuum luminosity measurements from SDSS spectra and miniJPAS photospectra. Triangles, squares, and stars indicate the continuum luminosity at $\lambda=5100$, 3000, and $1350\angstrom$, respectively. Blue and orange symbols display the results for sources with $r<20$ and $20<r<21$, error bars denote $1\sigma$-equivalent uncertainties, and the dashed line indicates a 1:1~relation between SES and SEP measurements.}
    \label{fig:lum}
\end{figure}

\subsection{Continuum luminosity}
\label{sec:results_continuum}

In Fig.~\ref{fig:lum}, we display continuum luminosity measurements from SDSS spectra and miniJPAS photospectra for the 54 quasars in the validation sample. Triangles, squares, and stars indicate the continuum luminosity at $\lambda=5100$, 3000, and $1350\angstrom$, respectively, and blue and orange symbols display the results for sources with $r<20$ and $20<r<21$, and the dashed line indicates a 1:1~relation between SES and SEP measurements. Error bars denote $1\sigma$-equivalent uncertainties from spectral and photometric decomposition. As we can see, these uncertainties do not capture the dispersion of measurements between SDSS spectra and miniJPAS photospectra; this is likely because error bars do not attempt to capture the impact of variability on the results, which is an important source on uncertainty in the comparison between spectroscopic and photometric measurements (see Sect. \ref{sec:data_sample}).

We find a small systematic difference between photometric and spectroscopic measurements: the mean and standard deviation of their difference are -0.11 and 0.11 dex for $\lambda L_\lambda(5100\angstrom)$, -0.07 and 0.11 dex for $\lambda L_\lambda(3000\angstrom)$, and -0.07 and 0.12 dex for $\lambda L_\lambda(1350\angstrom)$. We check that this bias is sensitive to the photometric correction used to convert aperture to total magnitudes (see Sect. \ref{sec:data_sample}), and that it disappears by slightly decreasing such a correction. Nonetheless, it is crucial to keep in mind that a constant bias has negligible impact on the computation of SMBH masses because the recalibration of virial coefficients absorbs it (see Sect. \ref{sec:model_masses}).


\begin{figure*}
    \centering
    \includegraphics[width=0.32\textwidth]{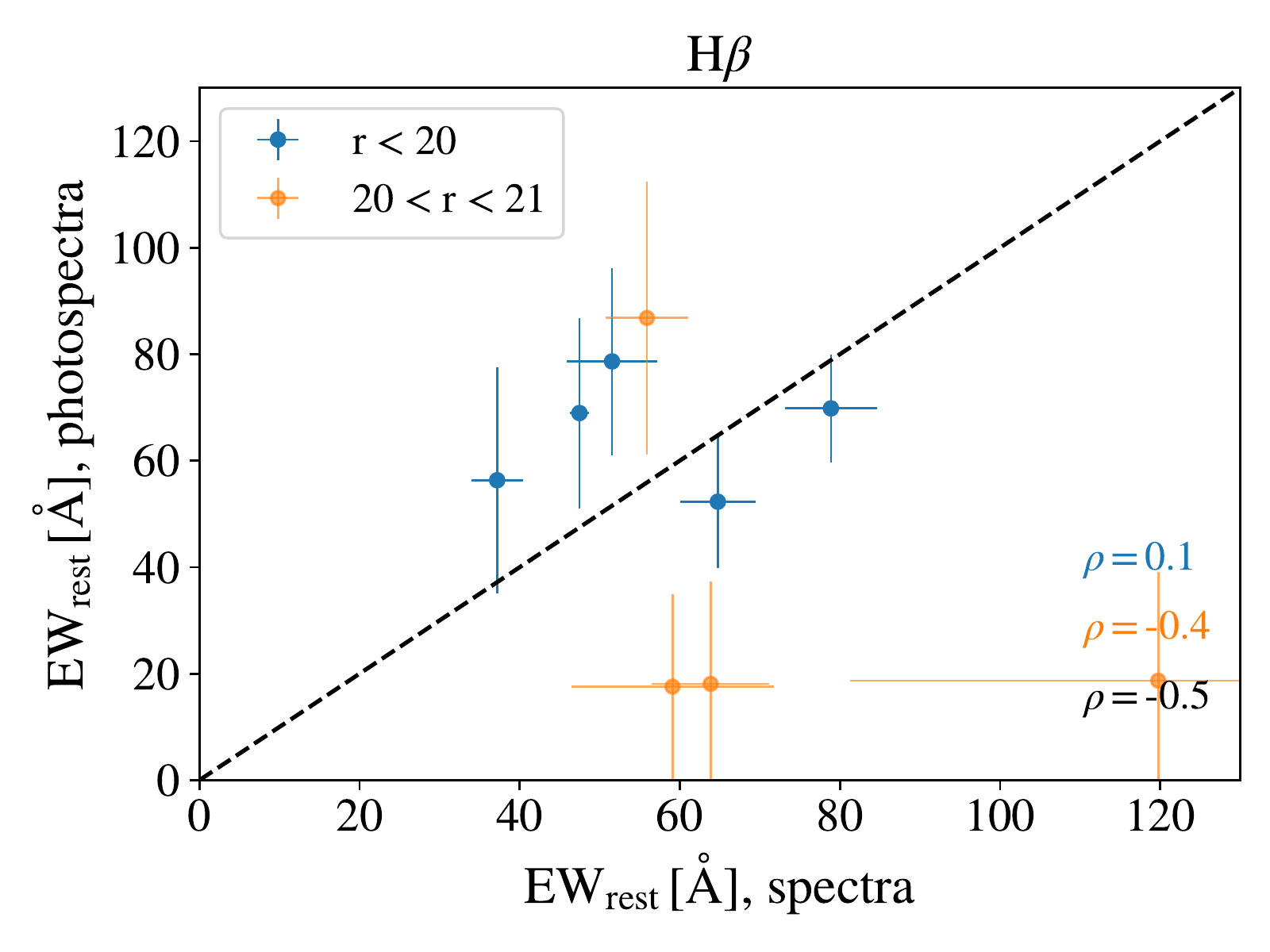}
    \includegraphics[width=0.32\textwidth]{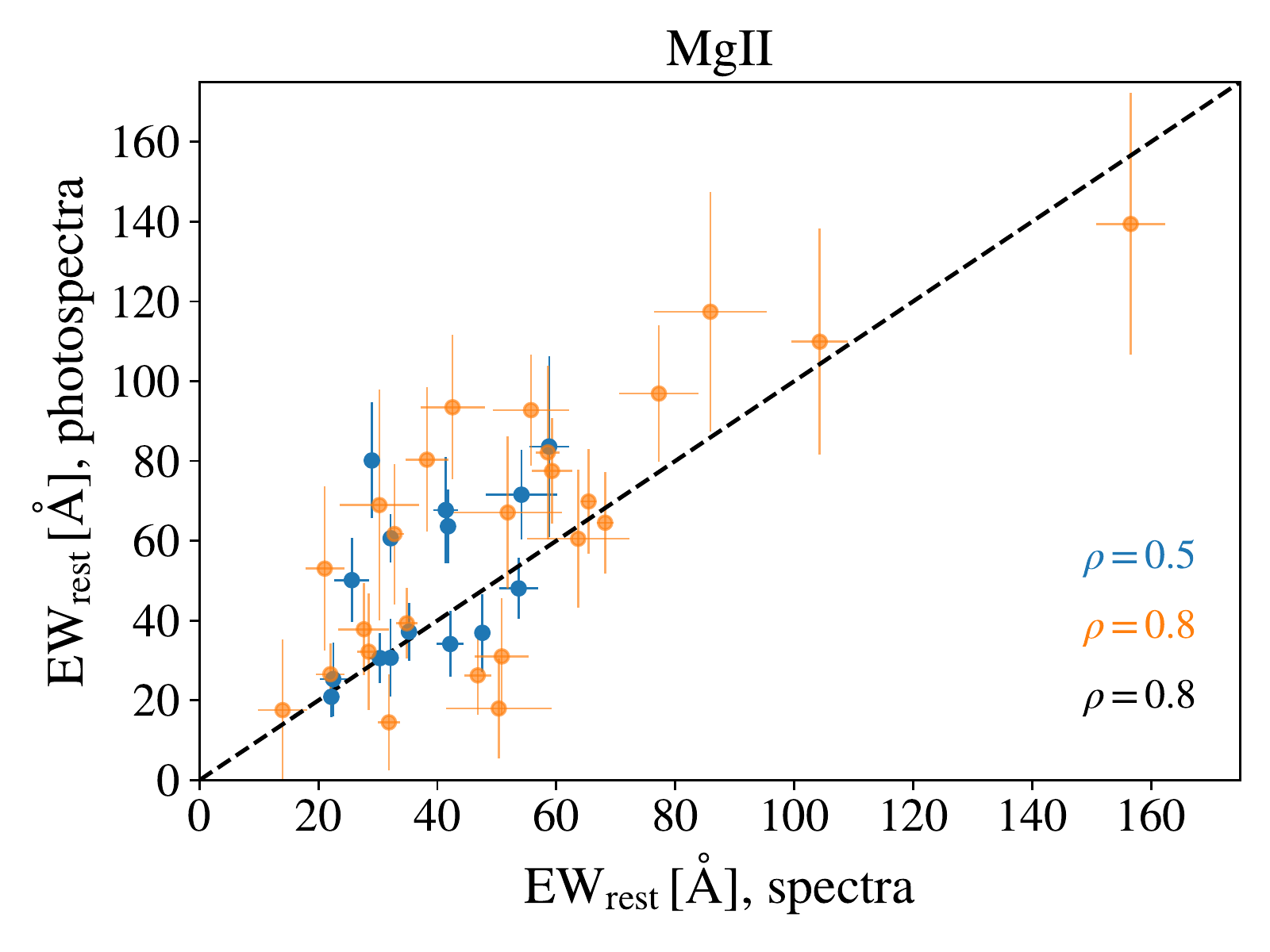}
    \includegraphics[width=0.32\textwidth]{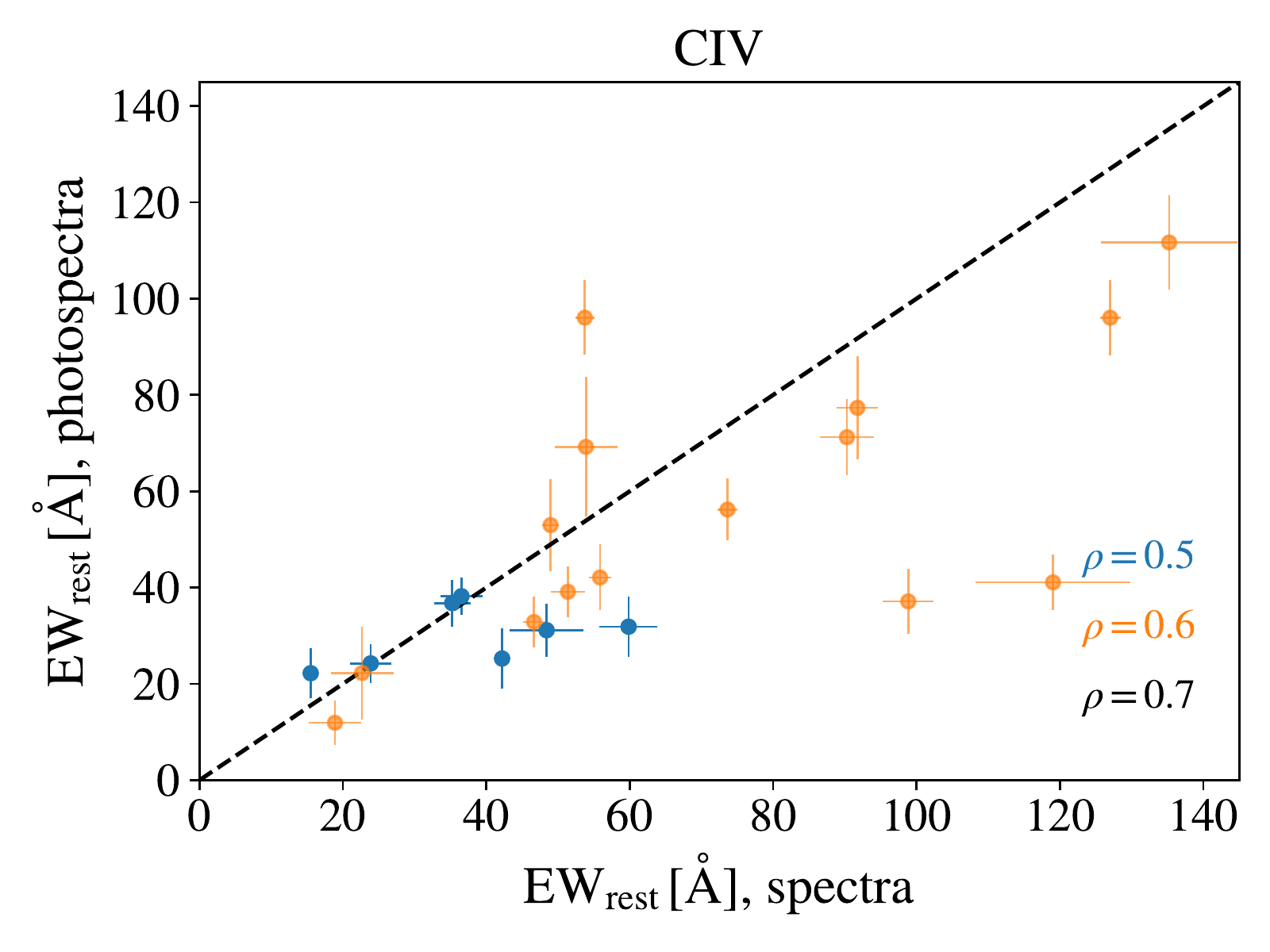}
    
    \includegraphics[width=0.32\textwidth]{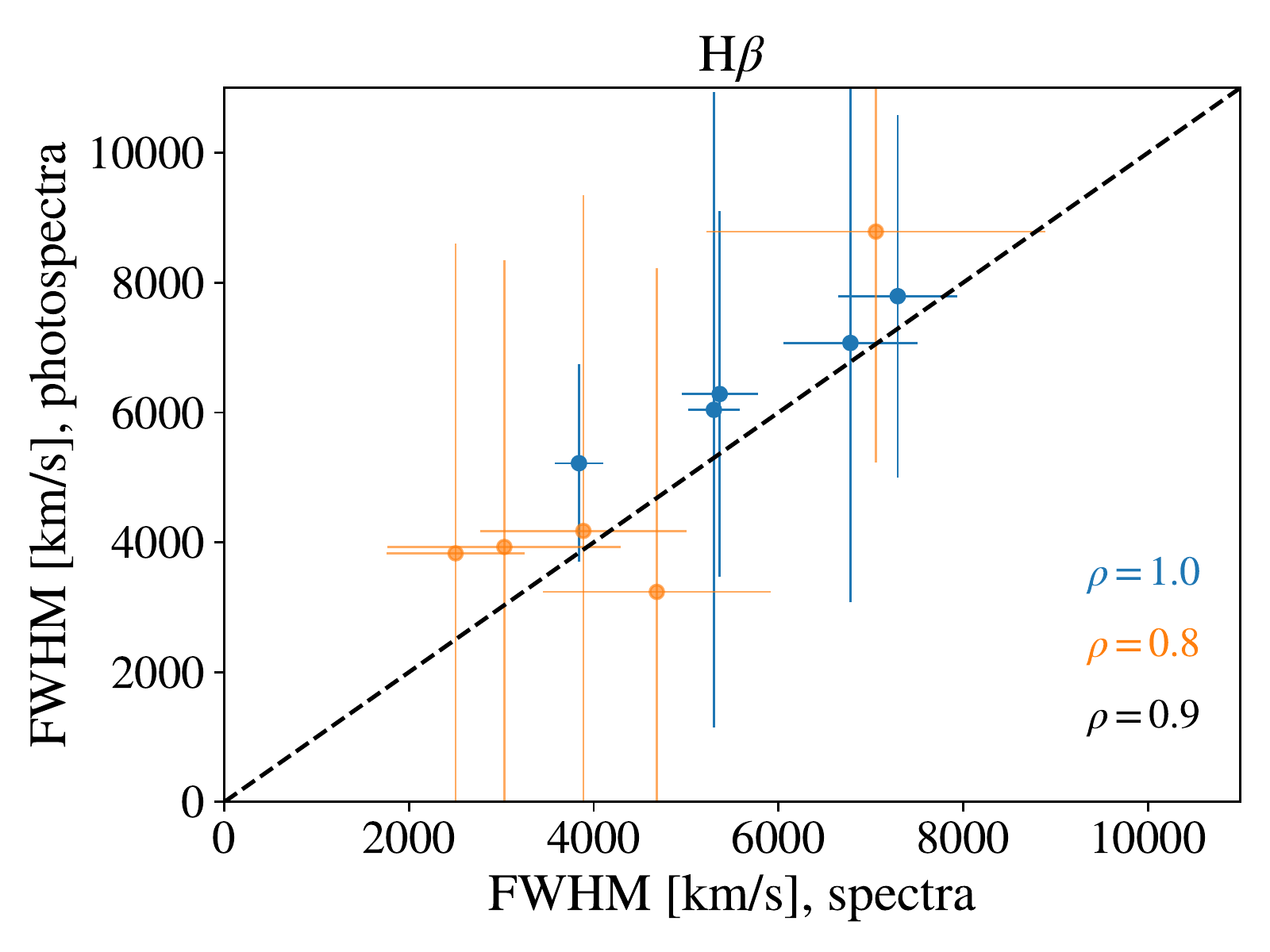}
    \includegraphics[width=0.32\textwidth]{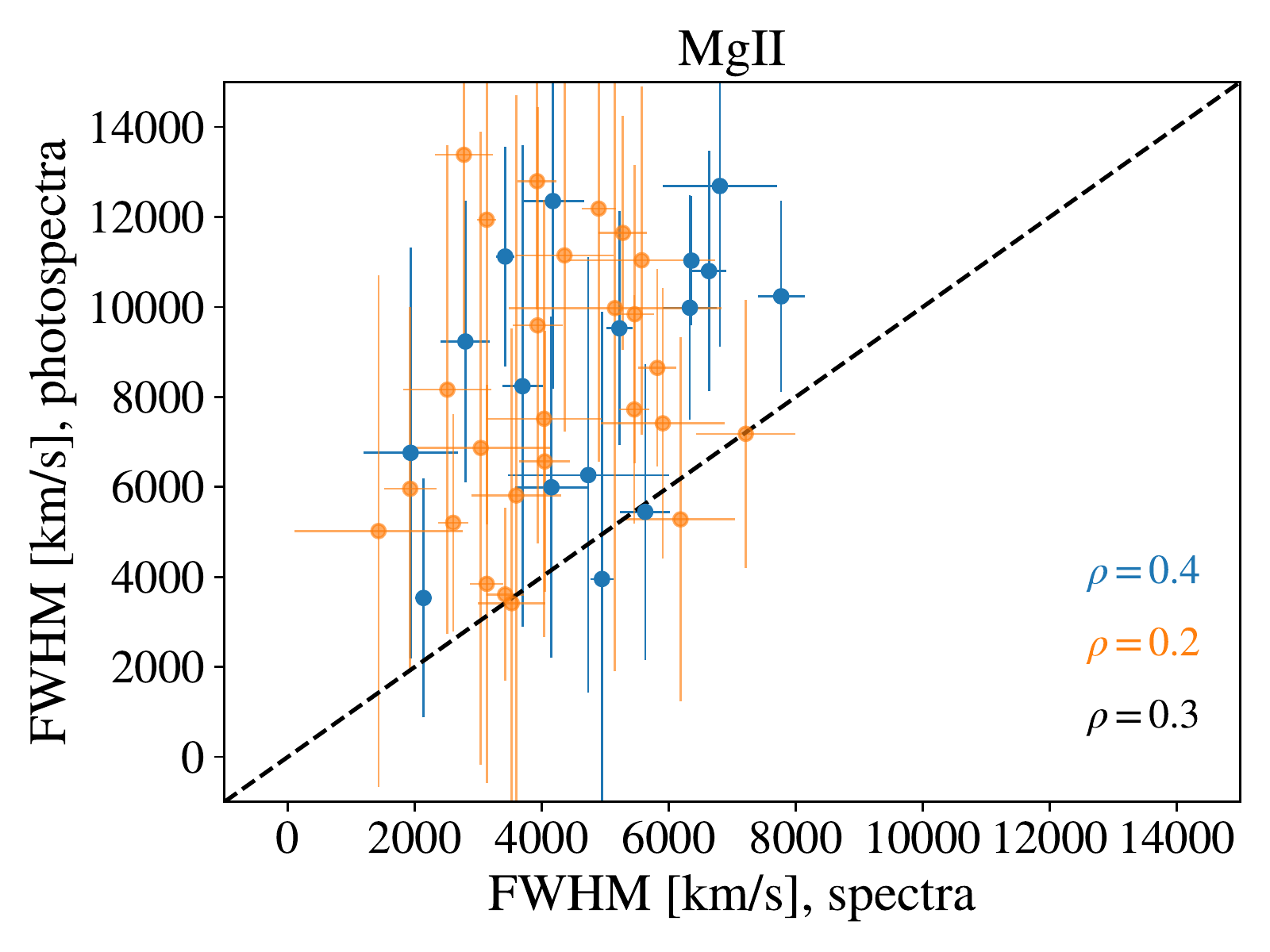}
    \includegraphics[width=0.32\textwidth]{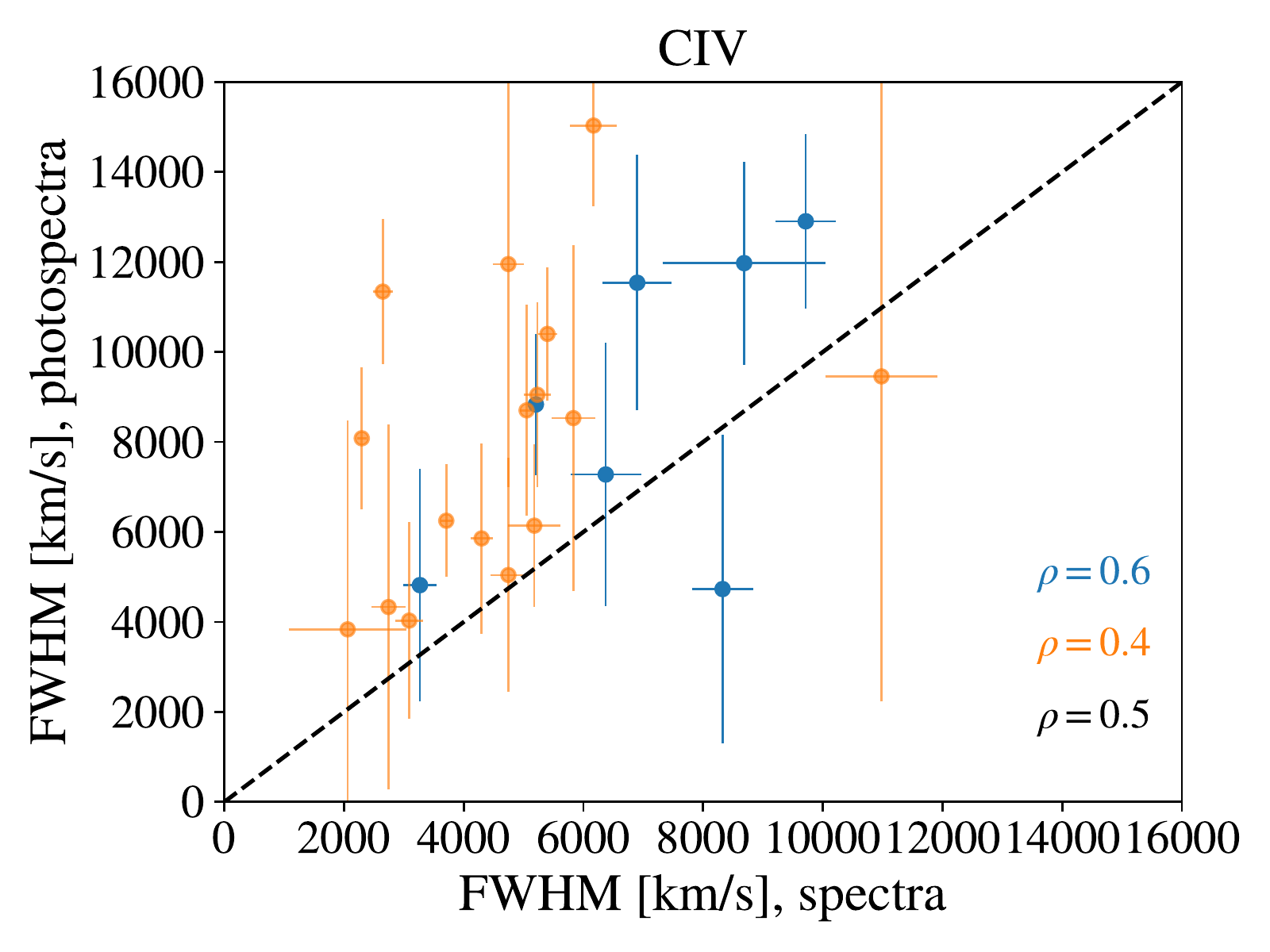}
    
    \caption{EW (top row) and FWHM (bottom row) measurements from SDSS spectra and miniJPAS photospectra. The left, middle, and right panels show the results for \hbeta, \mgii, and \civ, respectively. Dashed lines indicate a 1:1~relation between photometric and spectroscopic measurements. The symbol $\rho$ indicates the Pearson correlation coefficient between spectroscopic and photometric measurements.}
    \label{fig:lines}
\end{figure*}

\subsection{Emission line properties}
\label{sec:results_lines}

In Fig.~\ref{fig:lines}, we display EW (top row) and FWHM (bottom row) measurements from SDSS spectra and miniJPAS photospectra for the 54 quasars in the validation sample. The left, middle, and right panels show the results for \hbeta, \mgii, and \civ, respectively. These lines are broader than $1500\,\kms$ for all sources, and thus we expect accurate FWHM measurements from miniJPAS photospectra (see Sect. \ref{sec:model_mock}). The symbol $\rho$ indicates the Pearson correlation coefficient between spectroscopic and photometric measurements. As we can see, EWs and FWHMs present much larger error bars than continuum luminosity measurements, reflecting the complexity of extracting line properties. The median precision of EW and FWHM measurements from photometric data is 16 and 30\%, respectively, and from spectroscopic data is 6 and 8\%. Therefore, at fixed $r$-band apparent magnitude, spectroscopic measurements from SDSS are between two to four times more precise than photometric measurements from miniJPAS. We note that these error bars do not attempt to capture the impact of variability on the comparison between spectroscopic and photometric measurements (see Sect. \ref{sec:data_sample}).

We find that the Person correlation coefficient between spectroscopic and photometric FWHM measurements is almost unity for \hbeta, close to $\rho=0.5$ for \civ, and approximately $\rho=0.3$ for \mgii. The modest value of the correlation coefficient for some lines is likely caused by the combination of the limited S/N of observations quasar variability. Nonetheless, the agreement between emission line properties from spectroscopic and photometric measurements is substantial: we find that the mean and standard deviation of the difference for EWs are 0.15 and 0.36 dex for \hbeta, -0.07 and 0.19 dex for \mgii, and 0.09 and 0.17 dex for \civ, and for FWHM -0.06 and 0.09 dex for \hbeta, -0.28 and 0.19 dex for \mgii, and -0.20 and 0.18 dex for \civ. We thus find a systematic difference between FWHM measurements from SDSS spectra and miniJPAS photospectra, possibly because we use a single component to fit line profiles while \citet{rakshit2020_SpectralPropertiesquasars} does so using multiple components (see Sect. \ref{sec:model_lines}). We note that differences of the same order are found between widths measured by spectral decomposition methods using a single and multiple components \citep[e.g.][]{shen2011_CatalogQuasarProperties, rakshit2020_SpectralPropertiesquasars}. However, it is essential to notice that a constant bias has a negligible impact on the computation of SMBH masses because the recalibration of virial coefficients absorbs it (see Sect. \ref{sec:model_masses}).


\begin{figure*}
    \centering
    \includegraphics[width=0.85\textwidth]{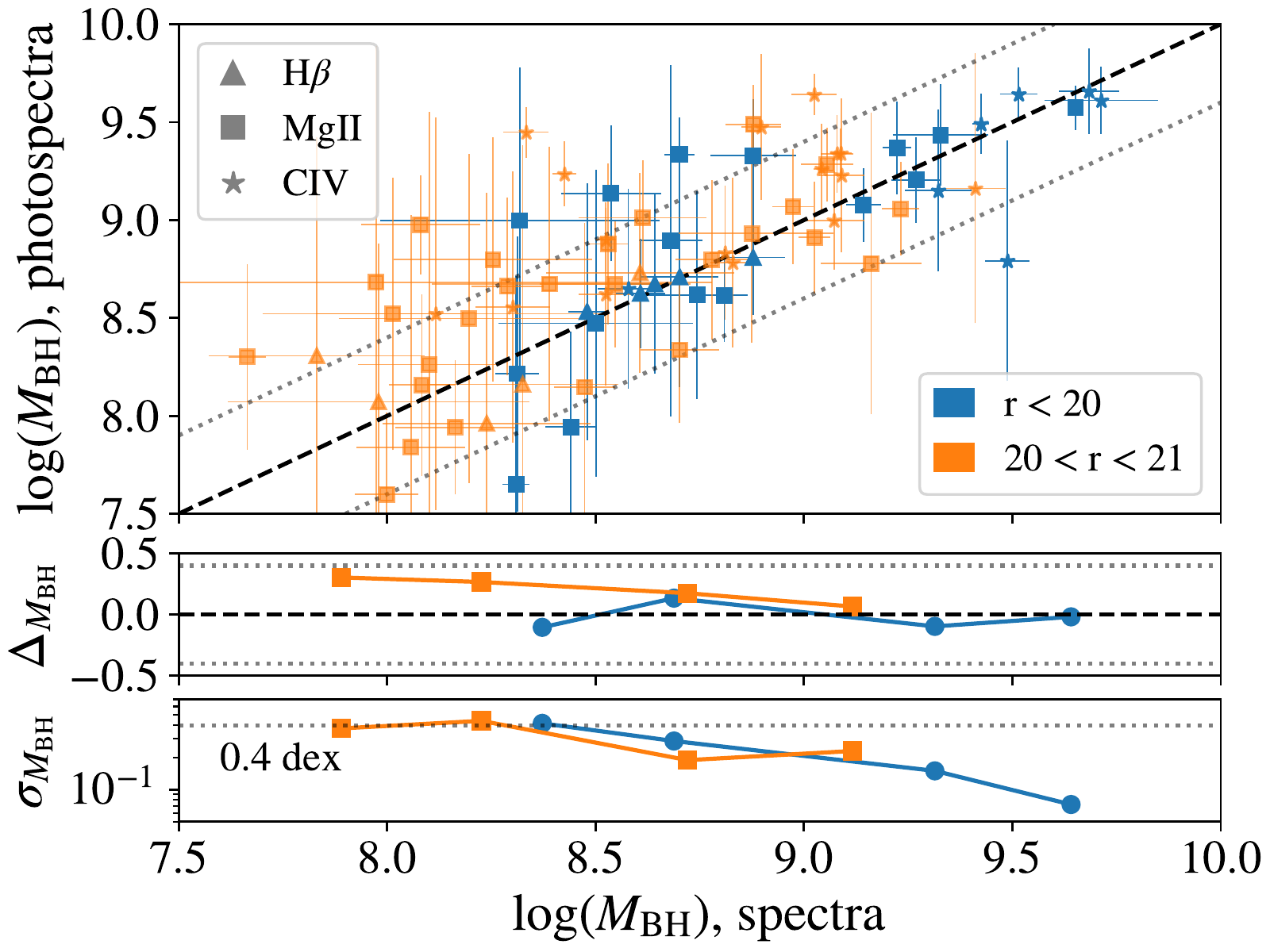}
    \caption{SMBH mass measurements from SDSS spectra and miniJPAS photospectra. The middle and bottom sub-panels display the mean and standard deviation of the logarithmic difference between spectroscopic and photometric measurements, respectively, the dashed line indicates a 1:1~relation between these measurements, and dotted lines are displaced 0.4 dex from such a relation. As we can see, there is a remarkable agreement between SES and SEP masses, with most measurements less than $1\sigma$ away from the 1:1~relation.}
    \label{fig:bmh}
\end{figure*}

\begin{figure}
    \centering
    \includegraphics[width=\columnwidth]{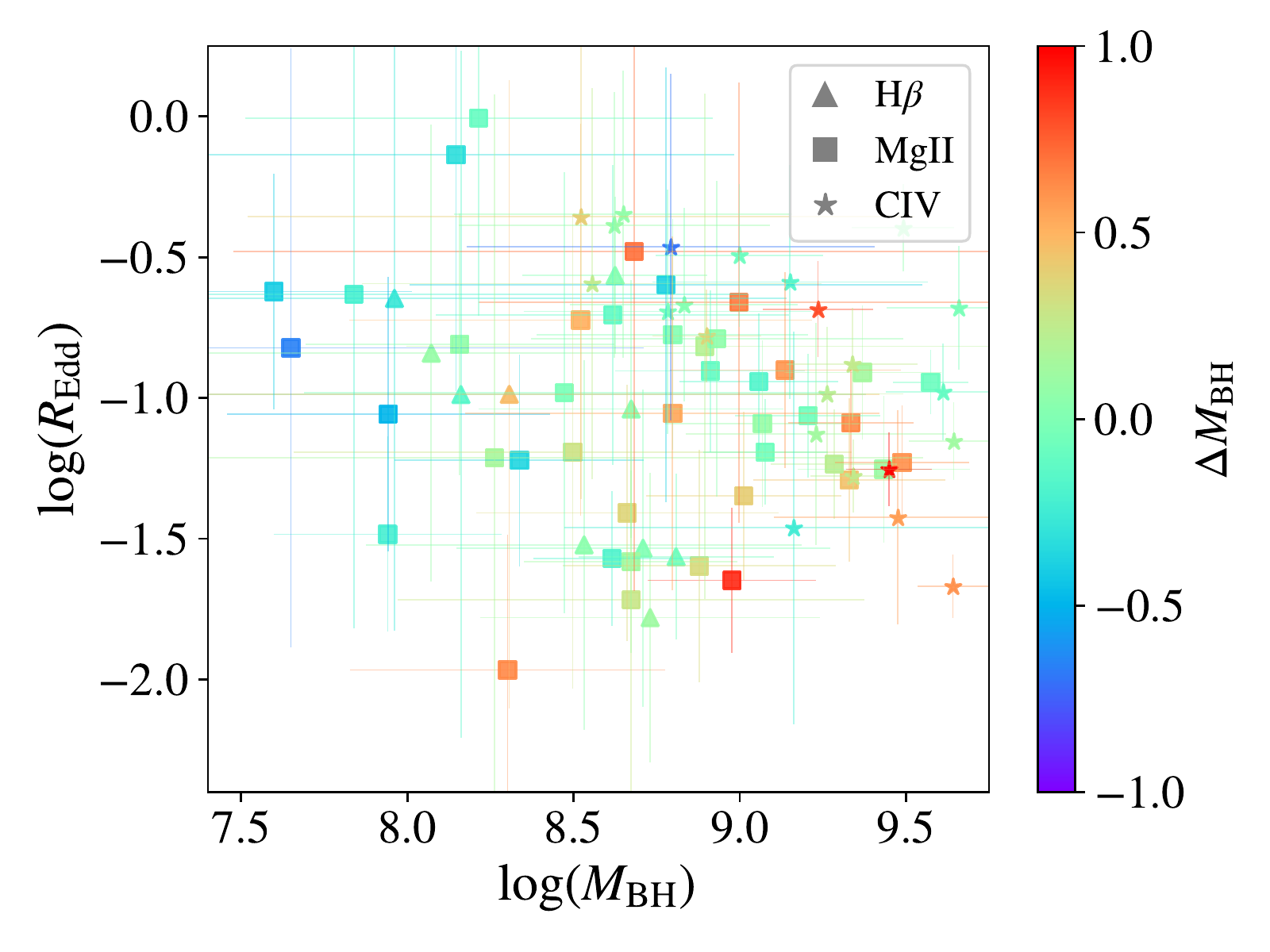}
    \caption{Performance of SEP as a function of Eddington ratio. Symbols indicate results from miniJPAS photospectra, and their colour denotes the logarithmic difference between mass measurements from SDSS spectra and miniJPAS photospectra. As we can see, the mass difference shows no dependence upon the Eddington ratio.}
    \label{fig:bhm_redd}
\end{figure}

\subsection{SMBH virial masses}
\label{sec:results_masses}

Before computing SMBH virial masses, we recalibrated the virial coefficients of Eq.~\ref{eq:mass_estimator} to correct for possible systematic differences between SES and SEP measurements due to the different methodologies to compute continuum luminosities and emission line properties. Following the approach explained in Sect. \ref{sec:model_masses}, we find $A=0.876\pm0.005$ and $B=0.263\pm0.013$ for \hbeta, $A=0.292\pm0.017$ and $B=0.590\pm0.006$ for \mgii, and $A=0.894\pm0.125$ and $B=0.263\pm0.029$ for \civ. 

In Fig.~\ref{fig:bmh}, we compare different SMBH mass estimates for quasars in the validation sample: SEP masses from miniJPAS photospectra produced using the previous coefficients and SES masses from SDSS spectra generated using the coefficients quoted in Sect. \ref{sec:model_masses}. The middle and bottom sub-panels show the mean and standard deviation of the logarithmic difference between spectroscopic and photometric measurements, respectively. Horizontal error bars denote $1\sigma$ uncertainties for SES masses, which are computed via error propagation of Eq.~\ref{eq:mass_estimator} \citep[see][]{rakshit2020_SpectralPropertiesquasars}. We compute error bars for SEP masses following the previous approach for a fair comparison between spectroscopic and photometric error estimates. As we can see, the size of error bars decreases with the mass of the sources for both techniques, reflecting the increasingly larger precision of measurements for brighter sources, and at fixed apparent magnitude, for sources with broader lines (see Sect. \ref{sec:model_masses}). We note that error bars do not capture the impact of systematic uncertainties affecting single-epoch masses.

We find that SEP and SES masses display a remarkable agreement, with no systematic difference between these and most measurements less than $1\sigma$ away from the 1:1~relation. We can thus conclude that SEP delivers unbiased measurements of SMBH masses despite the limited spectral resolution of J--PAS photospectra. We also find that the standard deviation of the difference between spectroscopic and photometric measurements decreases with the mass of the sources: it ranges from 0.4 to 0.07 dex for SMBHs with masses from $\log(M_\mathrm{BH}/\Msun)\simeq8$ to $9.75$, respectively. This level of precision is in line with our forecasts from simulated miniJPAS observations (see Sect. \ref{sec:model_mock}), suggesting limited impact unresolved emission lines, host-galaxy contamination, and other effects on continuum luminosity and emission line properties (see Sects. \ref{sec:model_continuum} and \ref{sec:model_lines}).

It is a well-established observational result that black hole mass estimates show a residual dependence on the Eddington ratio \citep{du2016_SupermassiveBlackHoles, du2018_SupermassiveBlackHoles, grier2017_SloanDigitalSky, du2019_RadiusLuminosityRelationshipDepends, martinez-aldama2019_CanReverberationmeasuredQuasars, fonsecaalvarez2020_SloanDigitalSky, bonta2020_SloanDigitalSky}, which serves as a measure of the accretion rate of the system. Therefore, it is conceivable that the performance of SEP could depend upon the Eddington ratio,
\begin{equation}
    R_\mathrm{Edd}= L_\mathrm{bol}/L_\mathrm{Edd},
\end{equation}
where $L_\mathrm{Edd}=1.257\times10^{38}(M_\mathrm{BH}/\Msun)$ is the Eddington luminosity, while $L_\mathrm{bol}$ is the bolometric luminosity, which we estimate from the monochromatic luminosity at 5100, 3000, and $1350\angstrom$ using $L_\mathrm{bol}=9.26\,L_{5100}$, $5.15\,L_{3000}$, and $3.81\,L_{1350}$, respectively \citep{shen2011_CatalogQuasarProperties}. We note that the previous expressions can result in uncertainties as large as a factor of two \citep{richards2006_SpectralEnergyDistributions}; this is because precise estimation of bolometric luminosities requires observations of the spectral energy distribution from X-ray to radio.

In Fig.~\ref{fig:bhm_redd}, we display Eddington ratio measurements from miniJPAS photospectra as a function of SEP masses. The colour of the symbols indicates the logarithmic difference between mass measurements from SDSS spectra and miniJPAS photospectra. As we can see, there is no apparent dependence of the mass difference upon the Eddington ratio, letting us conclude that the performance of SEP is independent of the Eddington luminosity of the system.

Another important consideration is that SEP faces similar systematic uncertainties as SES, including the chosen measurement that characterises line widths, different assumptions involving spectral decomposition methods, host-galaxy contamination, broad absorption lines, outflows, residual dependence of SMBH masses on the Eddington ratio, uncertainties in the relation between continuum luminosity and BLR size, variations in the virial factor with the BLR geometry, and possible non-reverberating components of some emission lines \citep[for a recent detailed discussion about these effects see][]{bonta2020_SloanDigitalSky}. The aforementioned effects induce differences of the order of 0.4 dex between RM and SES masses, with little dependence on SMBH mass \citep[e.g.][]{grier2019_SloanDigitalSky, homayouni2020_SloanDigitalSkya, bonta2020_SloanDigitalSky}. If we assume that the impact of the previous systematics on SES and SEP is of the same order, we can compare the precision of these techniques measuring SMBH masses. For SMBHs with masses of the order of $\log(M_\mathrm{BH}/\Msun)=8$, the standard deviation of the difference between SES and SEP masses is of the same order as the impact of systematic uncertainties, suggesting that SEP masses are $\sqrt{2}\simeq1.4$ times less precise than SES masses. Systematic uncertainties progressively dominate the error budget for SMBH with larger masses; for sources with $\log(M_\mathrm{BH}/\Msun)\simeq9.0$ and 9.5, SES masses are only 12 and 3\% more precise than SEP masses, respectively. We can thus conclude that SEP yields SMBH virial masses with only mildly lower precision than SES for the majority of sources in the validation sample. However, a more thorough study of the accuracy of SEP requires sources with both RM and SEP masses.


\section{Conclusions}
\label{sec:conclusions}

Precise measurements of SMBH masses are crucial for characterising the properties of the SMBH population, understanding the links between SMBHs and their host galaxies, and using quasars as large-scale structure tracers for multiple cosmological applications. In this work, we develop a novel approach for measuring SMBH virial masses from single-epoch, narrow-band photometric observations. We summarise our main findings as follows:

\begin{itemize}
    \item In Sect. \ref{sec:model} we describe a Bayesian-based approach for measuring continuum luminosities and emission line properties from narrow-band photometric data. We validate our methodology using simulated J--PAS observations, finding that they can deliver accurate continuum luminosities for sources brighter than $r=21$ and unbiased FWHM measurements for lines broader than $\simeq1500\,\kms$. For the kinds of quasars we consider here, this value translates into a minimum SMBH mass of approximately $\log(M_\mathrm{BH}/\Msun)=8$.
    
    \item In Sect. \ref{sec:results} we characterise the performance of our methodology using 54 SDSS quasars observed by the miniJPAS survey, a proof-of-concept project of the J--PAS collaboration covering ${\approx}1\,\mathrm{deg}^2$ of the northern sky using the 56 J--PAS narrow-band filters. We find that the standard deviation of the difference between SES measurements from SDSS and SEP measurements from miniJPAS is approximately 0.1 and 0.2 dex for continuum luminosities and FWHMs, respectively. However, we caution that quasar variability is a significant source of uncertainty for this comparison.
    
    \item In Fig.~\ref{fig:bmh} we compare SES masses from SDSS and SEP masses from miniJPAS, finding that both are compatible within error bars. We also show that the standard deviation of the difference between spectroscopic and photometric measurements ranges from 0.4 to 0.07 dex for masses from $\log(M_\mathrm{BH}/\Msun)\simeq8$ to $9.75$, respectively. Reverberation mapping studies show that SES masses are affected by systematic uncertainties of the order of 0.4 dex; given that SES and SEP face similar systematics, we can conclude that SEP yields SMBH virial masses with only mildly lower precision than SES for the majority of sources in the validation sample.
\end{itemize}

Throughout this work, we have focused on characterising the precision SEP for the J--PAS survey, which will soon start observing thousands of square degrees of the northern sky without applying any source preselection other than the photometric depth in the detection band. Therefore, SEP has the potential to provide details on the physical properties of new types of quasar populations that do not satisfy the preselection criteria of previous spectroscopic surveys. We have shown that our current technique delivers precise measurements only for sources brighter than $r=21$; to push our technique towards fainter magnitudes, we plan to measure SMBH virial masses from stacked photospectra of low signal-to-noise sources. Taken together, we expect J--PAS and SEP to be of paramount importance for completing our knowledge of SMBH demographics across cosmic time.

To estimate the performance of our methodology, we have relied upon SDSS sources with both SES and SEP measurements. Due to the limited precision of SES measurements, we would have rather used quasars with SEP and either RM or direct mass measurements; however, we did not find any source with these more precise mass measurements within the miniJPAS footprint. On the other hand, the J--PAS survey will soon observe quasars with these types of mass measurements, thereby enabling a better characterisation of the performance of SEP. Furthermore, we will have more sources available for statistical studies, which will enable a more precise assessment of the performance of our novel technique for sources with different Eddington ratios and redshifts.

Although our technique was developed to analyse J--PAS photospectra, SEP can also be used to compute SMBH masses for quasars observed by other surveys without substantial modifications. For instance, we could use this methodology to analyse data from multi-band surveys with enough spectral resolution to resolve the profile of broad emission lines, such as SHARDS and PAUS, and low-resolution spectroscopic surveys covering an extensive wavelength range, such as {\it Gaia} \citep{collaboration2012_MeasuringLargescalestructure} and the Spectro-Photometer for the History of the Universe, Epoch of Reionization and Ices Explorer (SPHEREx) survey \citep{dore2014_CosmologySPHEREXAllSky}.

\begin{acknowledgements}
      We thank the anonymous referee for useful comments and suggestions. This paper has gone through the internal review process of the J--PAS collaboration. We acknowledge useful discussion with Giorgio Calderone, Roberto Cid Fernandes and Rain Kipper and the feedback of Jifeng Liu. This work uses observations made with the JST/T250 telescope and PathFinder camera for the miniJPAS project at the Observatorio Astrof\'isico de Javalambre (OAJ) in Teruel, which is owned, managed, and operated by the Centro de Estudios de F\'isica del Cosmos de Arag\'on (CEFCA). This work made use of the following python packages: {\sc astropy} \citep{astropycollaboration2013_AstropyCommunityPython, astropycollaboration2018_AstropyProjectBuilding}, {\sc emcee} \citep{foremanmackey13}, {\sc ipython} \citep{perez2007_IPythonSystemInteractive}, {\sc matplotlib} \citep{hunter2007_Matplotlib2DGraphics}, {\sc mpi4py} \citep{dalcin2005_MPIPython, dalcin2008_MPIPythonPerformance, dalcin2011_ParallelDistributedcomputing, dalcin2021_Mpi4pyStatusUpdate}, {\sc numpy} \citep{harris2020_ArrayProgrammingNumPy}, {\sc pydoe2} \citep{rickardsjogrenanddanielsvensson2018_PyDOE2Experimentaldesign}, and {\sc scipy} \citep{virtanen2020_SciPyFundamentalalgorithms}. We acknowledge the OAJ Data Processing and Archiving Unit (UPAD) for reducing and calibrating miniJPAS data and the use of the Atlas EDR cluster at the Donostia International Physics Center (DIPC). Funding for the J--PAS Project has been provided by the Governments of Spain and Arag\'on through the Fondo de Inversi\'on de Teruel, European FEDER funding and the Spanish Ministry of Science, Innovation and Universities, and by the Brazilian agencies FINEP, FAPESP, FAPERJ and by the National Observatory of Brazil. Additional funding was also provided by the Tartu Observatory and by the J--PAS Chinese Astronomical Consortium. Funding for OAJ, UPAD, and CEFCA has been provided by the Governments of Spain and Arag\'on through the Fondo de Inversiones de Teruel; the Arag\'on Government through the Research Groups E96, E103, and E16\_17R; the Spanish Ministry of Science, Innovation and Universities (MCIU/AEI/FEDER, UE) with grant PGC2018-097585-B-C21; the Spanish Ministry of Economy and Competitiveness (MINECO/FEDER, UE) under AYA2015-66211-C2-1-P, AYA2015-66211-C2-2, AYA2012-30789, and ICTS-2009-14; and European FEDER funding (FCDD10-4E-867, FCDD13-4E-2685). J.C.M. and S.B. acknowledge financial support from Spanish Ministry of Science, Innovation, and Universities through the project PGC2018-097585-B-C22. B.T. acknowledges support from the Israel Science Foundation (grant number 1849/19).  C.Q. acknowledges support from Brazilian funding agencies FAPESP and CAPES. L.A.D.G. and R.G.D. acknowledge financial support from the State Agency for Research of the Spanish MCIU through the ``Center of Excellence Severo Ochoa'' award to the Instituto de Astrof\'isica de Andaluc\'ia (SEV-2017-0709), and R.G.D. also does it to the projects AYA2016-77846-P and PID2019-109067-GB100. C.H.M. acknowledges financial support from the Spanish Ministry of Science, Innovation, and Universities through the project PGC2018-097585-B-C2. A.E. acknowledges the financial support from the European Union - NextGenerationEU and the Spanish Ministry of Science and Innovation through the Recovery and Resilience Facility project J-CAVA. J. V. acknowledges the technical members of the UPAD for their invaluable work: Juan Castillo, Tamara Civera, Javier Hern\'andez, \'Angel L\'opez, Alberto Moreno and David Muniesa.
\end{acknowledgements}

%
  \bibliographystyle{aa} 
  \bibliography{biblio} 
%



\begin{appendix}

\section{Forward- versus backward-modelling observations}
\label{app:back}

\begin{figure}
    \centering
    \includegraphics[width=\columnwidth]{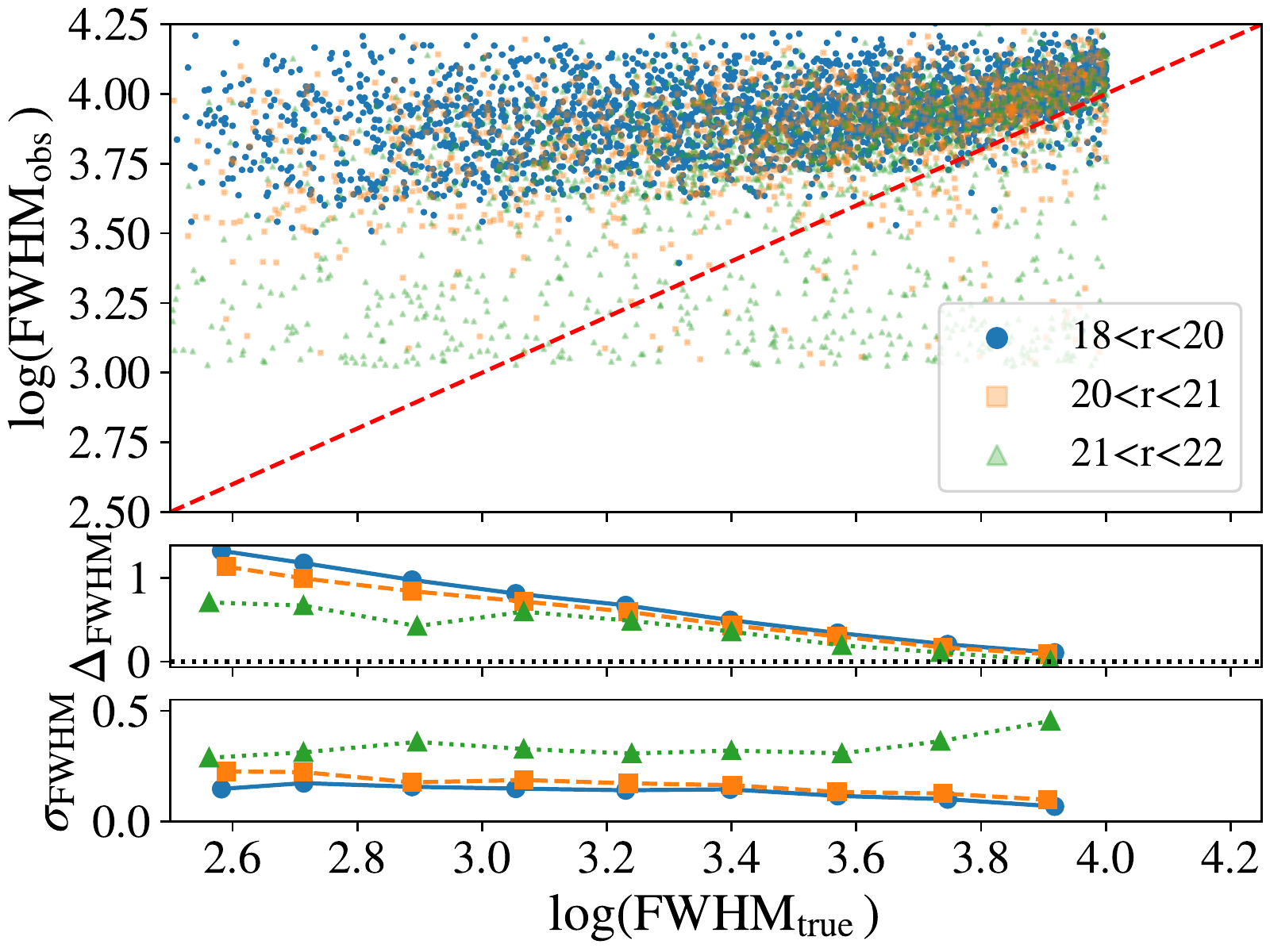}
    \caption{Backward-modelling measurements of \civ FWHMs from simulated J--PAS observations. This approach results in biased FWHM measurements even for lines as wide as $10\,000\,\kms$, while forward-modelling observations enable FWHMs of lines broader than $1500\,\kms$ in an unbiased fashion.}
    \label{fig:forecast_back}
\end{figure}

In this section we study if we can extract unbiased measurements of emission line widths by backward-modelling quasar observations.

To backward-modelling emission line widths, we first estimate the quasar continuum emission following the same approach as in Sect. \ref{sec:model_continuum}, and then we subtract it from observations to produce a line-only spectrum. After that, we identify the J--PAS bands in which the centre of the broad emission lines \hbeta, \mgii, and \civ fall; if one of these lines falls between two J--PAS bands, we select the band with the maximum flux. Finally, we compute the FWHM of each line by subtracting the wavelengths at which the continuum emission intercepts half the flux of the band at which the centre of the line falls. We note that we perform a linear interpolation of the flux between J--PAS bands to improve the resolution of the results.

In Sect. \ref{sec:model_mock}, we use simulated J--PAS photospectra to study the impact of both the limited spectral resolution of J--PAS and photometric errors on SEP. In Fig.~\ref{fig:forecast_back}, we show the result of backward-modelling the \civ width from simulated J--PAS observations. We can readily see that this approach results in biased FWHM measurements even for emission lines as broad as $10\,000\,\kms$. In Fig.~\ref{fig:forecast}, we show that forward-modelling observations enables the FWHM of lines as narrow as $1500\,\kms$ to be measured in an unbiased fashion, thereby pushing the accuracy of the results by almost an order of magnitude relative to backward-modelling. We thus conclude that forward-modelling observations is crucial for unbiased estimation of emission line properties from narrow-band surveys.

\section{Impact of photometric redshift errors}
\label{app:errors}

\begin{figure}
    \centering
    \includegraphics[width=\columnwidth]{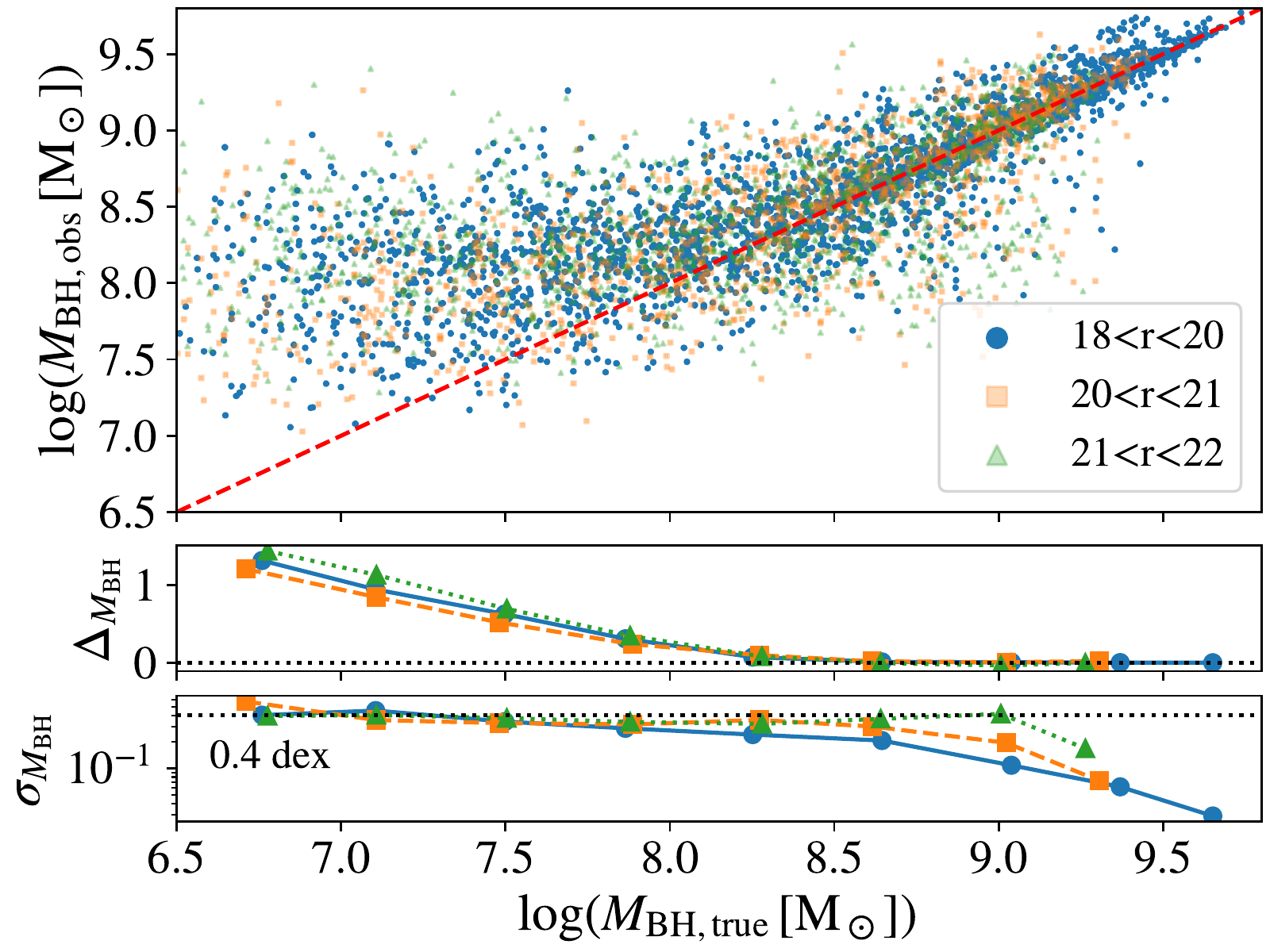}
    \caption{Dependence of SEP masses upon the level of photometric redshift errors expected for J--PAS quasars. Using simulated J--PAS observations, we find that the impact of these errors on SEP masses is negligible.}
    \label{fig:forecast_errors}
\end{figure}

In Sect. \ref{sec:results}, we use spectroscopic redshift estimates to conduct SEP measurements. The J--PAS survey will soon observe hundreds of thousands of quasars for which we will only have access to photometric redshift estimates. In this section we study the impact of photometric redshift errors on SEP measurements.

Forecasts for the J--PAS survey and preliminary results from the miniJPAS survey suggest that the precision of photometric redshifts for J--PAS quasars will be ${\approx}0.5\%$ \citep[][]{abramo12, chaves-montero2017_ELDARNewmethoda, bonoli2021_MiniJPASSurveypreview}. To study the impact of this level of uncertainties on SEP results, we first perturb the actual redshift of simulated J--PAS observations (see Sect. \ref{sec:model_mock}) according to a Gaussian of width $\sigma_z=0.005(1+z)$, and then we apply our methodology. In Fig.~\ref{fig:forecast_errors}, we show the precision of SEP masses measured from this sample. By comparing the results shown in this figure and the lower-right panel of Fig.~\ref{fig:forecast}, we can readily see that the impact of photometric redshift errors on SEP masses is negligible for the level of errors expected for J--PAS.

It is important to note that the previous approach does not account for the possibility of redshift outliers, that is, sources with photometric redshift estimate very far from their actual redshift primarily due to low S/N observations and line confusion \citep[e.g.][]{chaves-montero2017_ELDARNewmethoda}. We expect minimal impact of the first type of outliers because we only analyse sources brighter than $r=21$. On the other hand, outliers caused by line confusion are more problematic because these can be brighter than $r=21$, and our code will return precise measurements of the misclassified line. We will carry out a more detailed study about this source of uncertainty in future works.

\end{appendix}
\end{document}